\documentclass[12pt]{article}
\usepackage{graphicx}
\input{epsf.sty}
\def\mydate{12 July 2003}
\def\ignore#1{{}}

\let\oldtheequation=\theequation
\def\doteqs#1{\setcounter{equation}{0}
             \def\theequation{{#1}.\oldtheequation}}
\newcounter{sxn}
\def\sx#1{\addtocounter{sxn}{1} \vskip 1.cm  \goodbreak
\noindent{\large\bf\leftline{\thesxn.~~#1}} \nobreak \vskip -.6cm}
\def\sxn#1{\sx{#1} \doteqs{\thesxn}}

\newcounter{axn}
\def\ax#1{\addtocounter{axn}{1} \bigskip\medskip\goodbreak 
  \noindent{\large\bf {\Alph{axn}.~~#1}} \nobreak \medskip}

\def\axn#1{\ax{#1} \doteqs{\Alph{axn}}}

\date{}

\newdimen\mybaselineskip
\renewcommand{\baselinestretch}{1.25}

\newcommand{\beeq}{\begin{equation}}
\newcommand{\eneq}{\end{equation}}
\newcommand{\beqn}{\begin{eqnarray}}
\newcommand{\eeqn}{\end{eqnarray}}

\def\mybig{\displaystyle \strut }
\def\mbig{\displaystyle }
\def\dd{\partial}
\def\la{\raise.16ex\hbox{$\langle$}\lower.16ex\hbox{}  }
\def\ra{\, \raise.16ex\hbox{$\rangle$}\lower.16ex\hbox{} }
\def\go{\rightarrow}

\def\onehalf{ \hbox{${1\over 2}$} }

\def\Tr{{\rm Tr \,}}
\def\tr{{\rm tr \,}}
\def\eff{{\rm eff}}
\def\H{{\cal H}}
\def\G{{\cal G}}
\def\sym{{\rm sym}}
\def\SM{{\rm SM}}
\def\diag{{\rm diag ~}}
\def\BC{{\rm BC}}

\def\ep{\epsilon}
\def\psibar{ \psi \kern-.65em\raise.6em\hbox{$-$} }
\def\psibarl{ \psi \kern-.65em\raise.6em\hbox{$-$} \lower.6em\hbox{} }

\def\myfrac#1#2{{\mybig #1\over \mybig #2}}

\begin{document}
\thispagestyle{empty}

\baselineskip=12pt

{\small \noindent 22 January 2003  \hfill OU-HET 424/2002}

{\small \noindent  Corrected.  \mydate\hfill  hep-ph/0212035}

\baselineskip=40pt plus 1pt minus 1pt

\vskip 2.5cm

\begin{center}

{\LARGE \bf Dynamical Rearrangement of Gauge Symmetry on
the Orbifold $S^1/Z_2$}\\

\vspace{1.5cm}
\baselineskip=20pt plus 1pt minus 1pt

{\bf  Naoyuki Haba$^{1}$, Masatomi Harada$^{2}$, Yutaka Hosotani$^{2}$}\\
{\bf  and Yoshiharu Kawamura$^{3}$}\\
\vspace{.3cm}
$^1${\small \it Faculty of Engneering, Mie University,
Tsu, Mie, 514-8507, Japan} \\
$^2${\small \it Department of Physics, Osaka University,
Toyonaka, Osaka 560-0043, Japan}\\
$^3${\small \it Department of Physics, Shinshu University,
Matsumoto, Nagano 390-8621, Japan}\\
\end{center}

\vskip 2.cm
\baselineskip=20pt plus 1pt minus 1pt

\begin{abstract}
Gauge theory defined on the orbifold $M^4 \times (S^1/Z_2)$ is
investigated from the viewpoint of the Hosotani mechanism.
Rearrangement of  gauge symmetry takes place due to the dynamics
of Wilson line phases.  The physical symmetry
of the theory, in general, differs from the symmetry of the boundary
conditions.  Several sets of  boundary conditions having distinct
symmetry can be related  by gauge transformations, belonging to the same
equivalence class.  The Hosotani mechanism guarantees the same physics
in all theories in one equivalence class.   Examples are
presented in the $SU(5)$ theory. Zero modes of the extra-dimensional
components, $A_y$,  of gauge fields acquire masses by radiative
corrections.   In the nonsupersymmetric $SU(5)$ model the presence of bulk
fermions leads to the spontaneous breaking of color $SU(3)$.  In the
supersymmetric  model with  Scherk-Schwarz SUSY breaking  zero modes of
 $A_y$'s acquire masses of order of the SUSY breaking.
\end{abstract}


\newpage


\newpage

\sxn{Introduction}

Gauge theory defined in more than four dimensions have many attractive
features.  Interactions at low energies may be truely unified and
some of the distinct fields in four dimensions can be integrated
in a single multiplet in higher dimensions.  Higgs fields could be 
a part of  gauge fields.  Furthermore topology and structure of
extra-dimensional space  provide new ways of breaking symmetries,
accounting for, at the same time, the  hierarchy problem.

Higher dimensional gauge theory has long history.  It has been discussed
in the context of the Kaluza-Klein unification of gravity, gauge interactions,
and others.\cite{KK}  With the invent of the string theory higher 
dimensional theory,
which had been just curiosity of theorists till then,  has become important
ingredient and necessity in the present paradigm for the ultimate theory.
The string theory is consistently defined only in ten dimensions.\cite{string}
What we are left with after compactification of extra six-dimensional space
is a low energy theory of gravity and gauge interactions in four dimensions.
Without good understanding of the compactification mechanism
one cannot pin down which gauge theory to result at low energies.
The compactification may take place in several steps.  It is possible
that an effective gauge theory in five dimensions emerges
at an energy scale between the electroweak scale and the Planck scale.
We need to know how such higher dimensional theory reduces to the established
four-dimensional standard theory based on the gauge group
$SU(3) \times SU(2) \times U(1)$.

In this paper we shall consider five dimensional gauge theory, though  the
idea  and analysis  can be easily generalized to higher dimensions.
There are a couple of possibilities for the topology of the fifth
dimension.  One choice is a circle, $S^1$, which gives a multiply
connected manifold without a singularity.  On a multiply connected
manifold boundary conditions imposed on fields affect symmetry of the
theory. Scherk and Schwarz were the first to introduce such a twisted boundary
condition to break supersymmetry.\cite{SS,SS2}  In gauge theory there 
appear new degrees of freedom whose dynamics spontaneously break or restore
symmetry of the theory.\cite{YH1}-\cite{Takenaga}  Hosotani showed
that the dynamics of Wilson lines, which become physical degrees of freedom
on a multiply connected manifold and  parametrise degenerate
vacua at the tree level, lift  the degeneracy of vacua.
When the effective potential assumes the absolute minimum at a nontrivial
configuration of the Wilson line phases, the gauge symmetry can be
spontaneously broken or enhanced.  
Further theories with different boundary
conditions with different symmetries can be connected by the dynamics of the
Wilson lines, therefore falling under one equivalence class with the same
physics.  With all these interesting behavior, however, it is very difficult
to construct a realistic model. The main reason lies in the fact that a
simple manifold does not accommodate chiral fermions in four dimensions,
unless nontrivial topology is assigned to extra-dimensional space with
resulting complexity.

Major advance has been made recently.  Five dimensional spacetime may not
be a smooth manifold.  Instead it may be an orbifold such as $S^1/Z_2$ with
 four-dimensional hypersurfaces, branes, at the boundaries.
Such orbifolds naturally emerge in the string theory.\cite{orb}  Depending 
on how
the compactification proceeds $\lq$matter' fields such as gauge fields, Higgs
fields, and fermion fields may live only on the branes, or may live in the
bulk five-dimensional spacetime.  Randall and Sundrum showed, supposing that
only gravity lives in the bulk, that the gauge hierarchy problem can be
solved without fine tuning on $S^1/Z_2$.\cite{RS}  Kawamura showed that if 
the gauge
fields and Higgs fields live in the bulk in the $SU(5)$ model, the
triplet-doublet splitting problem is naturally solved on $S^1/(Z_2\times
Z_2')$.\cite{YK1,YK2,YK3}  Since then extensive investigation has been
made for constructing realistic grand unified models on 
orbifolds.\cite{Hall1}-\cite{orbGUTs4}

One aspect which has not been understood well is the role of the dynamics of
the Wilson line phases left over on orbifolds.  Recently Kubo, Lim and
Yamashita have analysed the $SU(3)$ model on $S^1/Z_2$ to find that the
vacuum shifts to a new one by quantum corrections, generating fermion
masses.\cite{Lim1}  This
effect brought by the Hosotani mechanism is generic in most of the gauge
theories on orbifolds.

We shall investigate   $SU(5)$ gauge theories on $S^1/Z_2$ with particular
attention on physics of the orbifold boundary condition and the dynamics
of the Wilson lines.  During the course of the investigation we have
encountered great amount of confusion in the literature on the issue.
We shall first give, in section 2,  general arguments for classifying the
equivalence classes of the boundary conditions, evaluating the effective
potential for the left-over Wilson lines, and determining the residual gauge
symmetry. In subsequent sections we give detailed analysis of $SU(5)$
models.  Wilson line degrees of freedom are pinned down and the effective
potential for those degrees of freedom is evaluated.  With typical orbifold
boundary conditions it is found in section 5 that the gauge
symmetry in the standard model, $G_\SM = SU(3) \times SU(2) \times U(1)$, is 
preserved only in the absence of bulk fermions.  In section 6 a thorough 
discussion is given about the dependence of theories on orbifold boundary 
conditions.  It is shown by explicit computation of the effective 
potential that so long as boundary conditions belong to the same equivalence
class, all theories yield the same physics content through the Hosotani 
mechanism.  A theory having $U(1) \times U(1) \times U(1)$ as the symmetry
of boundary conditions, for instance, actually has the physical symmetry
$G_\SM$ at the quantum level.  Such enhancement of the symmetry takes
place as a result of the dynamics of Wilson line phases.  Supersymmetric $SU(5)$ models
with soft SUSY breaking are analysed in Section 7.  It is shown that the
existence of more than one Higgs hypermultiplets in the bulk induces color $SU(3)$
breaking.  Extra-dimensional components of gauge fields corresponding to
Wilson line degrees of freedom acquire finite masses of order of the 
SUSY breaking scale by radiative corrections.  It is a generic feature
in those models that there appears a false vacuum which has higher energy
density than the true vacuum but is classically  stable.  In section 8
the quantum stability of the false vacuum is examined.  Surprisingly
we shall find there that the false vacuum is practically stable.
Technical details of computations are summarized in four appedices.  

We shall see in this paper how models with simple boundary
conditions have rich structure in the pattern of symmetry breaking and mass
generation.  It is a consequence of the dynamics of Wilson line phases.

\sxn{Orbifold  conditions and the Hosotani mechanism}

In this article we focus on a five-dimensional orbifold
$M^4 \times (S^1/Z_2)$ where $M^4$ is the four-dimensional Minkowski
spacetime.  The fifth dimension $S^1/Z_2$ is obtained by identifying
two points on $S^1$ by parity.  Let $x^\mu$ and $y$ be coordinates of
$M^4$ and $S^1$, respectively.   $S^1$ has a radius $R$ so that
a point $(x^\mu, y+2\pi R)$ is identified with a point $(x^\mu, y)$.
The orbifold $M^4 \times (S^1/Z_2)$ is obtained by further identifying
$(x^\mu, -y)$ and $(x^\mu, y)$.  The resultant fifth dimension is the
interval $0 \le y \le \pi R$.  As we shall see below, however, it is not
simply an interval.  It carries over the information on $S^1$.

\vskip .5cm
\leftline{\bf 2.1 Boundary conditions}

As a general principle the Lagrangian density has to be single-valued and
gauge invariant on $M^4 \times (S^1/Z_2)$.
In a gauge theory with a gauge group $G$ each field needs to return to
its original value after a loop translation along $S^1$ only up to
a global transformation of $G$.  We call it the $S^1$
boundary condition.  For a gauge field $A_M$
\beeq
U: ~ A_M (x, y + 2\pi R) = U A_M(x,y) U^\dagger 
\label{S1BC1}
\eneq
The $Z_2$-orbifolding is specified by parity matrices.
Around $y=0$
\beqn
P_0:
&&A_\mu (x, - y ) = P_0 A_\mu(x,y) P_0^\dagger \cr
\noalign{\kern 10pt}
&&A_y (x, - y ) = - P_0 A_y (x,y) P_0^\dagger
\label{OrbiBC1}
\eeqn
and around $y= \pi R$
\beqn
P_1:
&&A_\mu (x, \pi R - y ) = P_1 A_\mu(x,\pi R + y) P_1^\dagger \cr
\noalign{\kern 10pt}
&&A_y (x, \pi R - y ) = - P_1 A_y (x,\pi R + y) P_1^\dagger ~~.
\label{OrbiBC2}
\eeqn
To preserve the gauge invariance $A_y$ must have an opposite sign relative
to $A_\mu$ under these transformations.   As the repeated
$Z_2$-parity operation brings a field configuration back to the
original,  $P_0^2$ must be an element of the center of the group $G$.
By redefinition of $P_0$ one can suppose $P_0^2 = 1$, and therefore
$P_0^\dagger = P_0$.   The compatibility with the gauge invariance
demands that  $P_0$, with an appropriate phase factor,  is an element of $G$.   The
same conditions apply to $P_1$.

At this stage we observe that not all of $U$,  $P_0$ and $P_1$
are independent.  As a transformation $\pi R + y \go \pi R - y$ must be the
same as a transformation $\pi R + y \go - (\pi R + y) \go \pi R - y$,
it follows that
\beeq
U =  P_1 P_0   ~~.
\label{parity1}
\eneq
In case $\det (P_1 P_0) = -1$, $U$ need to be defined as
$e^{i\alpha} P_1 P_0$ such that $\det U=1$ for gauge groups, say,  $SU(N)$.
However, this phase factor does not affect the results below.
The definition  (\ref{parity1}) is adopted in the following discussions. 

For other fields it is more convenient to first specify the $Z_2$
parity conditions and then derive the $S^1$ condition.  For a 
scalar field 
\beqn
&&\hskip -1cm 
\phi(x,-y) = \pm T_\phi[P_0] \phi(x,y) \cr
\noalign{\kern 10pt}
&&\hskip -1cm 
\phi(x, \pi R -y) = \pm e^{i\pi\beta_\phi} T_\phi[P_1] \phi(x,\pi R + y) \cr
\noalign{\kern 10pt}
&&\hskip -1cm 
\phi (x, y + 2\pi R)  =  e^{i\pi \beta_\phi} T_\phi[U] \phi (x,y) ~~.
\label{OrbiBC3}
\eeqn
$T[U]$ represents an
appropriate representation matrix.  For instance, if $\phi$ belongs to the
fundamental or adjoint  representation of the group $G$, 
then $T_\phi [U] \phi$
is $U\phi$ or  $U \phi U^\dagger$, respectively.
In (\ref{OrbiBC3}) the relation $T_\phi[P_1] T_\phi[P_0]= T_\phi[U]$ has
been made use of.  There appears arbitrariness in
the sign provided the whole interaction terms in the Lagrangian
remain invariant.  As the repeated $Z_2$ parity operations must be
the identity operation, $e^{i\pi\beta_\phi}$ must be either $+1$ or $-1$, 
or equivalently to say,  $\beta_\phi$ has to be either 0 or 1. 

For Dirac fields defined in the bulk the gauge invariance of the kinetic
energy term  demands
\beqn
&&\hskip -1cm 
\psi(x,-y) = \pm   T_\psi[P_0] \gamma^5 \psi(x,y) \cr
\noalign{\kern 10pt}
&&\hskip -1cm 
\psi(x, \pi R -y) = \pm  e^{i\pi\beta_\psi} T_\psi[P_1]
  \gamma^5 \psi(x,\pi R + y)  \cr
\noalign{\kern 10pt}
&&\hskip -1cm 
\psi (x, y + 2\pi R)  = e^{i\pi \beta_\psi} T_\psi[U] \psi (x,y) ~~.
\label{OrbiBC4}
\eeqn
Just as for the scalar field, the phase factor
$e^{i\pi\beta_\psi}$ is restricted to be either $+1$ or $-1$, 
or equivalently  $\beta_\psi$ to be either 0 or 1.  $(\gamma^5)^2 = 1$
in our convention. 

One comment is in order.  In case there are
several multiplets, say,  $\phi^A$'s, in the same representation of the 
gauge group, there can be more general twisting in the flavor space.
The $S^1$ condition for $\phi$ in (\ref{OrbiBC3}) becomes, in general, 
\beeq
\phi^A (x, y + 2\pi R)  =  
{\big( e^{i\pi \beta M} \big)^A}_B T_\phi[U] \phi^B (x,y)
\label{S1BC2}
\eneq
where $M$ is a matrix in the flavor space.  If nontrivial
$Z_2$ parity is assigned in the flavor space and $M$ anti-commutes
with the $Z_2$ parity, then $\beta$ can take an arbitrary value.
Such an example naturally emerges in supersymmetric gauge 
theories.  In those theories a nontrivial
 $e^{i\pi \beta M}$ induces soft SUSY breaking.  We shall examine
supersymmetric $SU(5)$ models and come back to this point in section 7.

To summarize, the boundary conditions on $S^1/Z_2$ are specified with
$(P_0, P_1, U, \beta)$ and additional signs in
(\ref{OrbiBC3}) and (\ref{OrbiBC4}).  $(P_0, P_1, U)$ satisfies
$U=  P_1 P_0 $ and  $P_0^2 = P_1^2 = 1$.
We stress that $P_0$ and $P_1$ need not be diagonal in general.
A nontrivial example in the group $G=SU(2)$ is given by
\beeq
P_0 = \tau_3 ~~,~~
U = \pm \, e^{i(\alpha_1 \tau_1 + \alpha_2 \tau_2)}
~~,~~ P_1 = U P_0 ~~.
\label{BC1}
\eneq

In the literature the orbifold $S^1/(Z_2\times Z_2')$ has been
often considered,\cite{YK2} where the assignment of $P_0$ and $P_1$ is
given. It is equivalent to a gauge theory on $S^1/Z_2$ after rescaling of
the length of the interval.  When $P_0 \not= P_1$, $U=P_1 P_0 \not= 1$ so
that the $S^1$ boundary condition becomes nontrivial.

The orbifold $M^4 \times (S^1/Z_2)$ can be viewed as a manifold
with boundaries.\cite{HM}  At the boundaries $y=0$ and $y=\pi R$ appropriate
boundary conditions have to be imposed on fields such that
everything follows from the action principle.  As $P_0^2=P_1^2=1$,
eigenvalues of $P_0$ and $P_1$ are either $+1$ or $-1$, which implies
that with an appropriate basis chosen  fields obey
either Neumann or Dirichlet boundary  condition at each boundary.
Although this viewpoint is sometimes useful, there can arise
twist in the boundary conditions, i.e.\  the appropriate basis
on one boundary can be different from that on the other boundary.
Furthermore,  in gauge theory Wilson line degrees of freedom left over
under the orbifold conditions become dynamical and can lead to
dynamical alteration of the boundary conditions.  When the
Hosotani mechanism is operative, the orbifold viewpoint turns
more powerful and useful.

We would like to add a remark that nontrivial assignment of the
$Z_2$ parities provides a natural solution to the triplet-doublet mass
splitting problem and the chiral fermion problem.  Suppose that $[P_0, P_1]
= 0$ and
$U=P_1 P_0 = 1$.   Let $\Phi^+(x,y)$ and $\Phi^-(x,y)$ be fields with
$(P_0,P_1) = (1,1)$ and $(-1,-1)$, respectively.  They are expanded,
for $\beta_\Phi =0$,  as
\beqn
&& \hskip -1cm
\Phi^+(x,y) = {1\over \sqrt{\pi R}} ~ \phi^+_0(x) +
  \sqrt{{2\over \pi R} } \sum_{n=1}^{\infty} \phi^+_n (x)
        \cos \bigg( {ny\over R} \bigg)   ~~, \cr
\noalign{\kern 10pt}
&&\hskip -1cm
\Phi^- (x,y) = \sqrt{{2\over \pi R}}
    \sum_{n=1}^{\infty} \phi_n^- (x) \sin \bigg( {ny\over R}\bigg) ~~.
\label{expansion1}
\eeqn
The fields $\phi^{\pm}_n (x)$
acquire mass $n/R$ upon compactification.
Let $\Phi(x,y)$ be a multiplet in a symmetry
group.  The  symmetry reduction occurs at the classical level
unless all components
of $\Phi(x,y)$ have common $Z_2$ parities.
It is due to the absence of zero modes in the components with odd parity.
By use of this feature the triplet-doublet mass splitting in  the Higgs
multiplet is realized.  Four-dimensional theory with chiral fermions is also
constructed by projecting out their mirror fermions.

\vskip .5cm

\leftline{\bf 2.2 Residual gauge invariance of the boundary conditions}

Given the boundary conditions $(P_0, P_1, U, \beta)$, there
still remains the residual gauge invariance.  Recall that under a
gauge transformation $\Omega(x,y)$
\beqn
A_M &\to&  {A'}_M = \Omega A_M  \Omega^{\dagger}
   -  {i \over g}\Omega  \partial_M \Omega^{\dagger} ~~,  \cr
\phi &\to&
  \phi' = T_\phi [\Omega] \phi ~~,  \cr
\psi &\to& \psi' = T_\psi[\Omega] \psi  ~~.
\label{gauge1}
\eeqn
The new fields $A'_M$ satisfy, instead of
(\ref{S1BC1}) and (\ref{OrbiBC1}),
\beqn
A_M' (x, y + 2\pi R) &=& U' A_M'(x,y) U'^\dagger
   - {i\over g} U' \dd_M U'^\dagger \cr
\noalign{\kern 10pt}
\pmatrix{A_\mu' (x, - y) \cr  A_y' (x, - y) \cr}
&=& P_0' \pmatrix{ A_\mu' (x,y) \cr - A_y' (x,y) \cr} P_0'^\dagger
   - {i\over g} \,  P_0' \,
  \pmatrix{ \dd_\mu \cr - \dd_y \cr} P_0'^\dagger \cr
\noalign{\kern 10pt}
\pmatrix{A_\mu' (x,\pi R- y) \cr A_y' (x,\pi R- y) \cr}
&=& P_1' \pmatrix{A_\mu' (x,\pi R + y) \cr -A_y' (x,\pi R + y) \cr}
  P_1'^\dagger
   - {i\over g} \,  P_1' \,
\pmatrix{ \dd_\mu \cr -\dd_y\cr}  P_1'^\dagger
\label{BC2}
\eeqn
where
\beqn
&&\hskip -1cm
U' = \Omega(x,y+2\pi R) \,  U \, \Omega^\dagger (x,y) \cr
\noalign{\kern 5pt}
&&\hskip -1cm
P_0' = \Omega(x,-y) \, P_0 \, \Omega^\dagger (x,y) \cr
\noalign{\kern 5pt}
&&\hskip -1cm
P_1' = \Omega(x,\pi R -y) \, P_1 \, \Omega^\dagger (x,\pi R + y)  ~~.
\label{BC3}
\eeqn
Other fields $\phi'$ and $\psi'$ satisfy  relations similar to
(\ref{S1BC1}), (\ref{OrbiBC3}) and (\ref{OrbiBC4}) where
$(P_0, P_1, U)$ are replaced by  $(P_0', P_1', U')$.

The residual gauge invariance of the boundary conditions is given by gauge
transformations  which preserve the given boundary conditions, namely those
transformations which satisfy $U'=U$, $P_0'=P_0$, and $P_1'=P_1$;
\beqn
&&\hskip -1cm
\Omega(x,y+2\pi R) \,  U  =  U \, \Omega (x,y) \cr
\noalign{\kern 5pt}
&&\hskip -1cm
\Omega(x,-y) \, P_0  = P_0 \,  \Omega (x,y) \cr
\noalign{\kern 5pt}
&&\hskip -1cm
\Omega(x,\pi R -y) \, P_1 = P_1  \Omega (x,\pi R + y)  ~~.
\label{residual1}
\eeqn
We call the residual gauge invariance of the boundary conditions
the symmetry of the boundary conditions.   Explicit classification
in the $SU(2)$ model is given in Appendix A.
We remark that the symmetry of the boundary conditions in general differs
from the physical symmetry.  It may change at the quantum level by the
Hosotani mechanism.

Quite often we are interested in the symmetry at low energies, or more
precisely speaking, gauge invariance with $y$-independent gauge
transformation potential $\Omega= \Omega(x)$.  The low energy
symmetry of the boundary conditions is given by
\beqn
&&\hskip -1cm
\Omega(x) \,  U  =  U \, \Omega (x) \cr
\noalign{\kern 5pt}
&&\hskip -1cm
\Omega(x) \, P_0  = P_0 \,  \Omega (x) \cr
\noalign{\kern 5pt}
&&\hskip -1cm
\Omega(x) \, P_1 = P_1  \Omega (x)  ~~,
\label{residual2}
\eeqn
that is, the symmetry is generated by generators which commute with
$U$, $P_0$ and $P_1$.

\vskip .5cm
\leftline{\bf 2.3 Wilson line phases}

On a multiply connected manifold there appear new degrees of freedom
associated with a path-ordered integral along a noncontractible loop
$W =P \exp \Big\{ ig  \int_C dx^M A_M \Big\}$.   Non-integrable phases
of $W  U$ cannot be gauged away, and are called Wilson line
phases.\cite{YH2} Although constant Wilson line phases yield vanishing
field strengths at the classical level, they affect the spectrum of
excitations and  the symmetry of the theory at the quantum level.
The expectation values of the Wilson line phases are determined such that
the effective potential is minimized.  In lower dimensions quantum
fluctuations of the Wilson line phases become more dominant.  In quantum
electrodynamics on a circle, for instance, the dynamics of
the Wilson line phase lead to the $\theta$-vacuum.\cite{YH4,YH5}  In the
$2+1$ dimensional Chern-Simons theory on a torus the phases induce
the Heisenberg-Weyl algebra in the degenerate ground states as well.\cite{YH6}

On an orbifold $M^4 \times (S^1/Z_2)$ some of the Wilson line phases
on $M^4 \times S^1$ remain as physical degrees of freedom, depending on the
orbifold boundary conditions.  On  $M^4 \times (S^1/Z_2)$ Wilson line
phases correspond to $(x,y)$-independent modes of $A_y$.
It follows from  (\ref{OrbiBC1}) and (\ref{OrbiBC2}) that
$A_y =  - P_0 A_y P_0 = - P_1 A_y P_1$.  As $U = P_1 P_0$,
the relation $A_y = U A_y U^\dagger$  follows.

  Let us represent
\beeq
A_M = \sum_a A^a_M \, {\lambda^a \over 2} ~~,~~
\Tr \lambda^a \lambda^b = 2 \, \delta^{ab} ~~.
\label{decomposition1}
\eneq
A set,  $\cal G$, of the generators of the group $G$  is divided in two,
$\H_W$ and ${\cal G} - \H_W$;
\beeq
\H_W = \Bigg\{ ~{\lambda^a\over 2} ~;~
\{ \lambda^a, P_0 \} = \{ \lambda^a, P_1 \} =0 ~
\Bigg\} ~~.
\label{decomposition2}
\eneq
Wilson line phases on $M^4 \times (S^1/Z_2)$ are
$\{ \theta_a = g \pi R A_y^a ~,~ a \in \H_W \}$.

\vskip .5cm
\leftline{\bf 2.4 Equivalence classes of the boundary conditions}

Theory is specified with the boundary conditions.  Theories with
different boundary conditions can be equivalent in physics content.
The key observation is that in gauge theory one can always choose a gauge.
Physics should not depend on a gauge chosen.  Under (\ref{gauge1})
new fields satisfy boundary conditions (\ref{BC2}) and (\ref{BC3}).
If
\beqn
&&\hskip -1cm
\dd_M U' = 0 ~~,~~ \dd_M P_0' = 0 ~~,~~ \dd_M P_1' = 0 ~~, \cr
&&\hskip -1cm
P_0'^\dagger = P_0' ~~,~~ P_1'^\dagger = P_1' ~~,
\label{BC4}
\eeqn
then
\beeq
(U', P_0', P_1') \sim (U, P_0, P_1)
\label{equiv1}
\eneq
i.e.\  the two sets of the boundary conditions are equivalent.
It is easy to show that the relation $U' = P_1' P_0'$  is
maintained thanks to (\ref{BC4}).

The equivalence relation (\ref{equiv1})
defines equivalence classes of the boundary conditions.  We stress
that the boundary conditions indeed change under general gauge
transformations.  As an example, consider a $SU(2)$ gauge theory with
$(U, P_0, P_1)=(1,\tau_3, \tau_3)$.  Now make a gauge transformation
$\Omega = \exp \{ i(\alpha y/2 \pi R) \tau_1 \}$.  We find equivalence
\beeq
(1,\tau_3, \tau_3) \sim
(e^{i\alpha \tau_1}, \tau_3,  e^{i\alpha \tau_1} \tau_3 ) ~~.
\label{equiv2}
\eneq
The symmetry of the boundary conditions in one theory is also
different from that in the other.

\vskip .5cm
\leftline{\bf 2.5 The Hosotani mechanism}

Readers may be puzzled by the above result (\ref{equiv2}).  The two theories
with distinct symmetry of boundary conditions are equivalent to each other
in physics content.  How can  it be possible?  The equivalence is secured
by the  dynamics of the Wilson line phases.  It is a part of the Hosotani
mechanism.

Let us recall the Hosotani mechanism in gauge theories defined on multiply
connected manifolds.\cite{YH1,YH2}  It  consists of several parts.

\noindent  (i)  Wilson line phases along non-contractible loops become
physical degrees of freedom.  Once boundary conditions on fields are given,
Wilson line phases cannot be gauged away.  They
yield vanishing field strengths so that there appear degenerate vacua
at the classical level.

\noindent (ii) The degeneracy is lifted by quantum effects in general.
The effective potential $V_\eff$ for Wilson line phases $\theta$'s
acquires nontrivial dependence on $\theta$'s unless it is strictly forbidden
by such symmetry  as supersymmetry.  The physical vacuum is given by
the configuration $\theta$'s which minimizes $V_\eff$. (In two or three
dimensions significant quantum fluctuations appear around the minimum
of $V_\eff$.)

\noindent (iii) If the effective potential $V_\eff$ is minimized at a
nontrivial configuration of Wilson line phases, then the gauge symmetry
is spontaneously broken or restored by radiative corrections.   This part of
the mechanism is sometimes called the Wilson line symmetry breaking in the
literature.   Nonvanishing expectation values of the Wilson line phases
give masses to those gauge fields in lower dimensions whose gauge
symmetry is broken.  Some of matter fields also acquire masses.

\noindent (iv)  Nontrivial $V_\eff$ also implies that all extra-dimensional
components of gauge fields become massive. Their masses are given by
second derivatives of $V_\eff$ up to numerical constants.

\noindent (v) Two sets of boundary conditions for fields can be related
to each other by a boundary-condition-changing gauge transformation.
They are physically equivalent, even if the two sets have distinct
symmetry of the boundary conditions.  This defines equivalence classes of
the boundary conditions.  The effective potential
$V_\eff$ for Wilson line phases depends on the boundary conditions so that
the expectation values of the Wilson line phases depend on the
boundary conditions.  Physical symmetry of the theory is determined
by the combination of the boundary conditions and the expectation values
of the Wilson line phases.  Theories in the same equivalence class of the
boundary conditions have the same physical symmetry and physics content.

We need to discuss about the Hosotani mechanism on orbifolds.  The orbifold
conditions eliminate some or all of the Wilson line degrees of freedom.
Take a gauge theory on $M^4 \times (S^1/Z_2)$ discussed in the present
paper.  As described in (\ref{decomposition2}), surviving Wilson line
phases belong to the set $\H_W$.  Suppose that $\H_W$ is not
empty.  Then the  mechanism functions with no modification, provided
the equivalence classes of boundary conditions are defined as in subsection
2.4.

Let us spell out the  part (v) of the mechanism in gauge theory defined on
  $M^4 \times (S^1/Z_2)$.  One needs to first find physical symmetry of the
theory.  With the boundary conditions $(U, P_0, P_1)$  the effective
potential $V_\eff$ for the Wilson line phases is minimized at
\beqn
&&\hskip -1cm
\la  A_y \ra = {1\over 2\pi gR} \sum_{a\in {\H_W}} \theta_a \lambda^a \cr
\noalign{\kern 10pt}
&&\hskip -1cm
W \equiv \exp \Big\{ i 2\pi g R  \la A_y \ra \Big\}
= \exp \bigg\{i \sum_{a\in {\H_W}} \theta_a \lambda^a \bigg\} ~.
\label{Wphase1}
\eeqn
Now we make a gauge transformation given by
\beqn
&&\hskip -1cm
\Omega(y; \gamma) = S \bigg( { y\over 2 \pi R} + \gamma \bigg)   \cr
\noalign{\kern 10pt}
&&\hskip -1cm
S(z) = \exp \bigg\{ ~ i z
          \sum_{a\in {\H_W}} \theta_a \lambda^a \bigg\}
\label{gauge2}
\eeqn
where $\gamma$ is arbitrary.   Note $W = S(1)$.  In the new gauge
\beeq
\la A_y' \ra = \Omega \la  A_y \ra \Omega^{\dagger}
   -  {i \over g}\Omega  \partial_y \Omega^{\dagger}  = 0 ~,
\label{gauge3}
\eneq
i.e.\ the effective potential is minimized at the vanishing gauge
potentials.  However, the boundary conditions change;
\beqn
&&\hskip -1cm
P_0^\sym \equiv P_0' = \Omega(-y; \gamma) P_0 \Omega^\dagger(y; \gamma)
= S(\gamma) ~  P_0~  S(\gamma)^\dagger
\cr
\noalign{\kern 5pt}
&&\hskip -1cm
P_1^\sym \equiv
P_1' = \Omega(\pi R -y ; \gamma) P_1 \Omega^\dagger(\pi R + y; \gamma)
=S(\gamma )  ~W P_1 ~S(\gamma)^\dagger \cr
\noalign{\kern 5pt}
&&\hskip -1cm
U^\sym \equiv
U' = \Omega(y+ 2\pi R; \gamma) U \Omega^\dagger(y; \gamma)  =  W U ~~.
\label{BC5}
\eeqn
Here use of (\ref{parity1}) and (\ref{decomposition2}) has been made.
Therefore we have equivalence
\beeq
(U, P_0, P_1, \beta)
\sim (U^\sym, P_0^\sym, P_1^\sym, \beta ) ~.
\label{equiv3}
\eneq
Since the expectation values of the Wilson line phases vanish in the new
gauge, the physical symmetry of the theory is spanned by the generators
which commute with $(U^\sym, P_0^\sym, P_1^\sym)$;
\beeq
\H^\sym =  \Bigg\{ ~{\lambda^a\over 2} ~;~
[ \lambda^a, P_0^\sym ] = [ \lambda^a, P_1^\sym ] =0 ~
\Bigg\} ~~.
\label{decomposition3}
\eneq
The group, $H^\sym$, generated by $\H^\sym$ is the unbroken symmetry
of the theory.  Although $(P_0^\sym, P_1^\sym)$ depends on the parameter
$\gamma$, $H^\sym$ does not.

  Part (v) of the Hosotani mechanism presented above asserts that
if two sets of the boundary conditions are in the same equivalence
class, then the corresponding theories have the same $H^\sym$
among others.  We demonstrate it in the $SU(5)$ models in
Sections 5 and 6.

Dynamics of the Wilson line phases are at the core of the mechanism.
We would like to mention again that many attempts have been made
to utilize the mechanism on orbifolds to have coherent unified theories.
Kubo, Lim  and Yamashita have investigated the $SU(3)$ model.\cite{Lim1}
Further advance has been made by Gersdorff, Quiros and Riotto  to achieve
spontaneous supersymmetry breaking in the gauged supergravity
model.\cite{Quiros1}   In  the rest of
the paper we attempt to construct $SU(5)$ models on
$M^4 \times (S^1/Z_2)$ to incorporate natural solution to the
triplet-doublet splitting problem.

\sxn{Orbifold conditions in $SU(5)$ gauge theory}


As stressed in the introduction one of the attractive features of gauge
theories defined on orbifolds is that the hierarchy problem in the
conventional four-dimensional grand unified theory may be naturally
solved.  In particular,  the symmetry reduction by a non-trivial $Z_2$
parity  assignment ($P_0, P_1$)  offers a
powerful tool to construct a realistic grand unified model realizing the
triplet-doublet splitting naturally.\cite{YK1,YK2}

In this and subsequent sections we shall study five-dimensional  $SU(5)$
gauge theories with non-trivial $Z_2$ parity assignments to understand a
role of non-integrable Wilson line phases  on $M^4 \times (S^1/Z_2)$.
Our visible world is assumed to be one of the four-dimensional
hypersurfaces at the boundaries of the five-dimensional space-time.
For the moment we suppose that
gauge bosons $A_M(x, y)$ and some other fields live in  the bulk
five-dimensional spacetime and the
$SU(5)$ gauge symmetry is broken down to that of the standard model,
$G_\SM = SU(3) \times SU(2) \times U(1)$, by a non-trivial $Z_2$  parity
assignment.   The argument will be generalized in Section 6.

\ignore{
In subsection \ref{Z2}, we discuss $Z_2$ parity assignment,
gauge symmetry reduction and zero modes of gauge bosons.
We give a generic formula of effective potential in background field
method in \ref{eff-pot} and discuss the gauge invariance of effective
potential and its consequence in \ref{gauge-inv}. In subsection
\ref{SU(5)model} , we study a 5D $SU(5)$ model with a non-trivial
$Z_2$ parity assignment based on one-loop effective potential.
}


There are two types of $Z_2$ parity assignments which reduce
$SU(5)$ symmetry to $G_{SM}$  at the classical level, or equivalently,
have $G_{SM}$ as the symmetry of the boundary conditions.
They are
\beqn
{\rm Case~1}&&
P_0 = \pm \mbox{diag}(1,1,1,1,1) ~~,~~
P_1 = \pm \mbox{diag}(1,1,1,-1,-1) ~~, \cr
&& U =  \mbox{diag}(1,1,1,-1,-1) ~~,
\label{case1} \\
{\rm Case~2}&&
P_0 = \pm \mbox{diag}(1,1,1,-1,-1) ~~,~~
P_1 = \pm \mbox{diag}(1,1,1,-1,-1) ~~, \cr
&& U =  \mbox{diag}(1,1,1,1,1) ~~.
\label{case2}
\eeqn
The assignment in Case 1 has been employed in refs.\ \cite{YK2}
and \cite{YK3}.
Surviving  zero modes of gauge fields are $A_{\mu}^{a}(x)$ where
$a$ is the index of the generators of $G_{SM}$ of the standard model.
There are no zero modes for the fifth-dimensional component $A_y$ of the
gauge fields. The $SU(5)$ gauge symmetry is broken at the fixed point $y =
\pi R$ by the orbifold condition $P_1$.  There are no Wilson line phases,
i.e. $\H_W$ is empty.

The boundary conditions in Case 2, which have been  considered  in ref.\
\cite{YK1},  have the same symmetry, $G_{SM}$, of the
boundary conditions as in Case 1.  However, the physics content is quite
different.   There appear Wilson line phases.  $\H_W$ in
(\ref{decomposition2}) is
\beeq
{\rm Case~2~:} \quad
\H_W = \bigg\{ {\lambda^a\over 2} ~;~
a= 13 \sim 24 ~ \bigg\}
\label{Wphase2}
\eneq
where $\lambda^a$'s are defined in Appendix B.  Notice that $\H_W$ is
complementary to the set, $\G_\SM$,  of the generators of $G_\SM$;
$\G = \G_\SM + \H_W$.  The zero modes, namely the constant modes, of
$A_y^a$ ($a \in \H_W$) give Wilson line phases.

Dynamics of the Wilson line phases set in.  As discussed in the previous
section, the dynamics give $A_y^a$'s ($a \in \H_W$) finite masses at the
quantum level.  Furthermore, these $A_y^a$'s may develop nonvanishing
expectation values, depending on the matter content residing in the bulk
five-dimensional spacetime.  If that happens, the physical symmetry of the
theory is reduced, $SU(3)$ color of $G_\SM$ being broken.
In other words $\H^\sym \not= \G_\SM$.    One needs to
evaluate the effective potential $V_\eff$ for the Wilson line phases to
know if that happens.
If  $V_\eff$ is minimized at vanishing $A_y^a$'s ($a \in \H_W$), then
$G_\SM$  remains intact at the quantum level.  Otherwise we end up with a
theory with broken color, which cannot be accepted on phenomenological
grounds.

Various sets of the boundary conditions belong to the same equivalence
class as that in Case 2.  They have various symmetry of the boundary
conditions.  The set in Case 2 has $G_\SM$, whereas some others have
either $[SU(2)]^2  \times [U(1)]^2$, or
$SU(2)  \times [U(1)]^3$, or $ [U(1)]^3$.
Those sets are continuously connected by the Wilson lines, $\theta_a$'s.
The absolute minimum of the effective potential $V_\eff (\theta_a)$
determines the true vacuum and the physical symmetry of the theory
in this equivalence class.
In the rest of the paper we evaluate $V_\eff$ for various
matter content as well as the masses of $A_y^a$'s ($a \in \H_W$).
The structure of the equivalence class of the boundary conditions is
also clarified in due course.

\ignore{
At first sight, the existence of $A_y^{\hat{a}(0)}(x)$ induces
two serious problems on phenomenological grounds.
One is that it is fear that $A_y^{\hat{a}(0)}(x)$
get the vacuum expectation values (VEVs) radiatively
and it triggers to break the charge and/or color symmetry.
The other is that the theory predicts massless scalar  fields
$A_y^{\hat{a}(0)}(x)$  at tree level, which have not been found.
We conjecture that there is a minimum where the charge and/or color   is
unbroken beyond tree level and the $A_y^{\hat{a}(0)}(x)$ acquire a large
mass by the quantum correction there. We will check whether or not these
conjectures hold true by using one-loop effective potential in the
following subsections.
}

\sxn{Effective potential}


The effective potential can be evaluated as in gauge theory on multiply
connected manifolds.  The only necessary modification is to incorporate
the additional orbifold boundary conditions in the gauge fixing term.

Let us summarize the effective potential at one-loop level in the background
field method. The system is described by the following gauge fixed
Lagrangian density for non-Abelian gauge theory on D-dimensional space-time,
\begin{eqnarray}
&&\hskip -1cm
{\cal{L}} = {\cal{L}}_{gauge} + {\cal{L}}_{matter}  ~~,  \cr
\noalign{\kern 10pt}
&&\hskip -1cm
{\cal{L}}_{gauge}  = -{1 \over 2} \mbox{Tr} F_{MN}
F^{MN} - {1 \over \alpha} \mbox{Tr} F[A]^2  - \mbox{Tr}  \left(\bar{\eta}
{\delta F[A] \over \delta A_M} D^M \eta \right) ~~,  \cr
\noalign{\kern 10pt}
&&\hskip -1cm
{\cal{L}}_{matter}
= \bar{\psi} i \gamma_M D^M \psi + |D_M \phi|^2 - V[\phi, \psi] ~~,
\label{Lag}
\end{eqnarray}
where the second and third terms in ${\cal{L}}_{gauge}$ are the gauge-fixing
term with a gauge parameter  $\alpha$ to be chosen to be
$\alpha = 1$  and the ghost term, respectively.
$\psi$ and $\phi$ generically denote  Dirac fermion fields
and  complex scalar fields.  On orbifolds there cannot be bare
Dirac mass terms.
The covariant derivative is defined by $D_M \equiv \partial_M +  i g
T^a A_M^a$ where $T^a$ is an appropriate representation
matrix of the gauge group.

The  background field method is outlined as follows.

(1) We split the gauge field $A_M$ into the classical part $A_M^0$  and the
quantum part $A_M^q$. The $A_M^0$ is called the background field and  
is chosen so as to solve the classical equation of motion and satisfy the
gauge-fixing condition $F[A^0] = 0$. The $A_M^q$ is a variable of
integration in the path-integral  formalism.

(2) We choose the following gauge-fixing condition,
\begin{eqnarray}
F[A] = D_M(A^0) A^{qM} = \partial_M A^{qM} + ig [A_M^0, A^{qM}] = 0 ~~.
\label{gauge-fixing}
\end{eqnarray}
The covariant derivative $D_M(A^0)$ is often denoted by $D_M^0$ for short.
In this gauge there is the residual gauge invariance discussed in
section 2.2;
\begin{eqnarray}
A_M^{0} &\to&  {A'}_M^{0} =  \Omega A_M^{0} \Omega^{\dagger}
    -  {i \over g}\Omega \partial_M \Omega^{\dagger} ~~,  \cr
\noalign{\kern 5pt}
A_M^{q} &\to&  {A'}_M^{q} = \Omega A_M^{q} \Omega^{\dagger} ~~,  \cr
\noalign{\kern 5pt}
\psi~~ &\to&  \psi' = T_\psi [\Omega] \psi ~~, ~~~
  \phi \to \phi' = T_\phi[\Omega]
\phi ,
\label{gauge4}
\end{eqnarray}
where $\Omega$ is the gauge transformation matrix and
$T[\Omega]$ is an appropriate representation matrix of the gauge group.
The $A_M^q$ behaves as an adjoint matters.
$\Omega$ must be subject to (\ref{residual1}).

(3) Using the field equation for $A_M^0$
and the condition $F[A^0] = 0$,  we rewrite ${\cal{L}}_{gauge}$ to
obtain,  up to quadratic  terms in $A^q$,
\begin{eqnarray}
{\cal{L}}_{gauge} = - \mbox{Tr} A_M^{q} M_{MN}^{g} A^{Nq}
-  \mbox{Tr} \bar{\eta}  M^{gh} \eta
\label{Lag-gauge}
\end{eqnarray}
where $M_{MN}^{g}$ and $M^{gh}$ are defined by
\begin{eqnarray}
&&\hskip -1cm
M_{MN}^{g} = - \eta_{MN} D_{L}^{0} D^{0L} -4ig F_{MN}^{0} ~~, \cr
\noalign{\kern 5pt}
&&\hskip -1cm
M^{gh} = D_{L}^{0} D^{0L} ~~ .
\label{M}
\end{eqnarray}
Integrating out the quantum fields $A_M^q$, $\bar{\eta}$, $\eta$,  $\psi$ and
$\phi$, we obtain the one-loop effective potential for $A_M^0$;
\begin{eqnarray}
&&\hskip -1cm
V_\eff [A^0] = V_\eff [A^0]^{g+gh} + V_\eff [A^0]^{fermion}
+  V_\eff [A^0]^{scalar} ~~, \cr
\noalign{\kern 10pt}
&&\hskip -1cm
V_\eff [A^0]^{g+gh} = - (D-2) \,
{i \over 2} \, \Tr \ln  D_{L}^{0} D^{0L}  ~~, \cr
\noalign{\kern 10pt}
&&\hskip -1cm
V_\eff [A^0]^{fermion} = f(D){i \over 2}
  \, \Tr \ln  D_{L}^{0} D^{0L} ~~, ~~~ f(D) = 2^{[D/2]} ~~,\cr
\noalign{\kern 10pt}
&&\hskip -1cm
V_\eff [A^0]^{scalar} = - 2 \, {i \over 2}  \, \Tr \ln D_{L}^{0} D^{0L} ~~.
\label{Veff}
\end{eqnarray}
Here we have supposed that $F_{MN}^0 = 0$ and $\phi$-fields are massless.


The effective potential in the background field gauge has gauge invariance.
In this regard the dependence of the effective potential on the boundary
conditions has to be carefully treated.  In gauge theory on
  $M^{D-1} \times (S^1/Z_2)$,   $V_\eff [A^0]$ depends on
the boundary condition parameters
$( P_0, P_1, U, e^{i\beta})$ defined in Section 2.1.
\begin{eqnarray}
V_\eff [A^0] = V_\eff [A^0; P_0, P_1, U, \beta]  ~~.
\label{Veff2}
\end{eqnarray}

Now consider a boundary-condition-changing gauge transformation
introduced in Section 2.4 to define equivalence classes of the
boundary conditions.  A gauge potential $\Omega$ in (\ref{gauge1}) or
(\ref{gauge4}) must satisfy the condition (\ref{BC4});
\beeq
\dd_M U' = 0 ~~,~~ \dd_M P_0' = 0 ~~,~~ \dd_M P_1' = 0 ~~, ~~
P_0'^\dagger = P_0' ~~,~~ P_1'^\dagger = P_1' ~~,
\label{BC6}
\eneq
where
\beqn
&&\hskip -1cm
U' = \Omega(x,y+2\pi R) \,  U \, \Omega^\dagger (x,y) \cr
\noalign{\kern 5pt}
&&\hskip -1cm
P_0' = \Omega(x,-y) \, P_0 \, \Omega^\dagger (x,y) \cr
\noalign{\kern 5pt}
&&\hskip -1cm
P_1' = \Omega(x,\pi R -y) \, P_1 \, \Omega^\dagger (x,\pi R + y)  ~~.
\label{BC7}
\eeqn
The set $( P_0', P_1', U', e^{i\beta})$ is in the same equivalence class
as the set $( P_0, P_1, U, e^{i\beta})$.
The action is invariant under the gauge transformation
except the gauge-fixing term.
If the  relation
\begin{eqnarray}
D^{0M}(\partial_M \Omega^{\dagger} \Omega)
= \partial^M (\partial_M \Omega^{\dagger} \Omega) + i g[A^{0M},  \partial_M 
\Omega
\Omega^{\dagger}] = 0  ~~,
\label{BC8}
\end{eqnarray}
is satisfied, then the gauge fixing term is also invariant under the gauge
transformation as
\begin{eqnarray}
D^{M}(A^{0}) A_M = 0 \to D^{M}({A'}^{0}) A'_M = \Omega D^{M}(A^{0}) A_M
\Omega^{\dagger} = 0 ~~.
\label{gauge-inv-gf}
\end{eqnarray}
The entire action is gauge invariant under gauge transformations subject
to (\ref{BC6}) and (\ref{BC8}).
Hence the effective potential satisfies the relation
\begin{eqnarray}
V_\eff [A^0; P_0, P_1, U, \beta]
= V_\eff [A'^0; P_0', P_1', U', \beta] ~~.
\label{Veff3}
\end{eqnarray}

There are two special cases.  For transformations leaving $(P_0, P_1, U)$
unchanged, the relation (\ref{Veff3}) implies that
$V_\eff [A^0] = V_\eff [A'^0]$, i.e.\ $V_\eff [A^0]$ is a function
of invariant quantities under the symmetry of the boundary conditions.
In particular, it is invariant under global transformations satisfying
(\ref{residual2}).  Secondly, the relation  (\ref{Veff3}) applies to a
gauge transformation which brings $(P_0, P_1, U)$ to
$(P_0^\sym, P_1^\sym, U^\sym)$ defined in section 2.5.   As
(\ref{gauge2}) certainly satisfies the condition (\ref{BC8}) for
$A^0 = \la A \ra$ in the theory with $(P_0, P_1, U)$,
\beeq
V_\eff [\la A \ra ; P_0, P_1, U, \beta]
= V_\eff [A =0 ; P_0^\sym, P_1^\sym, U^\sym, \beta] ~~.
\label{Veff4}
\eneq
The set $(P_0^\sym, P_1^\sym, U^\sym)$ determines the  physical symmetry of
the theories in each equivalence class.

We apply the above results to Case 2, (\ref{case2}).
Configulations of interest are constant Wilson line phases having
$\la F_{MN} \ra = 0$.  Since $\H_W$ is given by (\ref{Wphase2}), we
parametrize $A^0$ in $V_\eff$ as
\beeq
A^0_y = {1 \over 2 g R} \pmatrix{
0 & \Theta \cr \Theta^{\dagger} & 0\cr}
\label{Ay}
\eneq
where $g$ is a 5D gauge coupling constant related to  4D one $g_4$  by
$g_4^2 = g^2/\pi R$.    $\Theta$ is a $3 \times 2$ matrix.
The set of the boundary conditions in Case 2 has the symmetry of the
boundary conditions $G_\SM$.  Under a global transformation in $G_\SM$
$\Theta$ is transformed as
\begin{eqnarray}
\Theta \to \Theta' = e^{i \alpha} \Omega_3 \Theta \Omega_2^{\dagger}
\label{gauge-trTheta}
\end{eqnarray}
where  $\Omega_3$, $\Omega_2$ and  $e^{i \alpha}$ are transformation
matrices of $SU(3)$, $SU(2)$, and $U(1)$, respectively.
$\Theta \Theta^{\dagger}$ is an $SU(2) \times U(1)$ invariant quantity
transforming as $\Theta \Theta^{\dagger} \to \Theta' {\Theta'}^{\dagger}  =
\Omega_3 \Theta \Theta^{\dagger}  \Omega_3^{\dagger}$,
whereas $\Theta^{\dagger} \Theta$ is an $SU(3) \times U(1)$ invariant
quantity transforming as
$\Theta^{\dagger} \Theta \to {\Theta'}^{\dagger} \Theta'
= \Omega_2  \Theta^{\dagger} \Theta \Omega_2^{\dagger}$.

Because the effective potential $V_\eff$ has $G_\SM$ invariance, we have
$V_\eff[\Theta] =V_\eff[\Theta']$.  As $\Theta$ is a $3 \times 2$ matrix,
there are only two invariants.  To see it, we first apply a global $G_\SM$
transformation to bring $\Theta$ in the form
\beeq
\Theta = \pmatrix{ \alpha & \gamma \cr  0 & \beta \cr 0 & 0\cr}
\label{Theta}
\eneq
where $\alpha$ and $\gamma$ are complex parameters and $\beta$ is a real
parameter. Using these parameters, $\Theta \Theta^{\dagger}$ and
$\Theta^{\dagger} \Theta$ are written as
\begin{eqnarray}
\Theta \Theta^{\dagger} = \left(
\begin{array}{ccc}
|\alpha|^2+|\gamma|^2 & \beta \gamma & 0 \\
\beta \gamma^* & \beta^2 & 0 \\
0 & 0 & 0
\end{array}\right)  ~~,~~
\Theta^{\dagger} \Theta = \left(
\begin{array}{cc}
|\alpha|^2 & \gamma \alpha^* \\
\gamma^* \alpha & \beta^2 + |\gamma|^2\\
\end{array}\right) ~~ .
\label{ThetaTheta}
\end{eqnarray}
The eigenvalues for $\Theta \Theta^{\dagger}$ are given by $0$,
  $\lambda_+$ and
$\lambda_-$ and those for $\Theta^{\dagger} \Theta$ are given by
  $\lambda_+$ and
$\lambda_-$. Here $\lambda_{\pm}$ are given by
\begin{eqnarray}
\lambda_{\pm} = {1 \over 2}
\left( \beta^2 + |\alpha|^2 + |\gamma|^2 \pm \sqrt{(\beta^2 + |\alpha|^2 +
|\gamma|^2)^2  - 4 |\alpha|^2 \beta^2}\right) .
\label{lambda}
\end{eqnarray}
$V_\eff$ is a function of invariants of $\Theta \Theta^{\dagger}$ and
$\Theta^{\dagger} \Theta$ as well.
Hence $V_\eff$ is regarded as a function of the two parameters $\lambda_{+}$
and  $\lambda_{-}$.
This implies that one can further simplify the form of $\Theta$ without
loss of generality.  We adopt  a simple choice that $\alpha = a$,
$\beta = b$  and $\gamma = 0$ where $a$ and $b$ are real.
In this case, $\lambda_{\pm}$ are given by $a^2$ and $b^2$.
In subsequent sections we evaluate $V_\eff$ for
\beeq
\Theta = \pmatrix{ a & 0 \cr  0 & b \cr 0 & 0\cr} ~~.
\label{Theta2}
\eneq
The resultant $V_\eff(a,b)$ should be interpreted as
$V_\eff(\sqrt{\lambda_+}, \sqrt{\lambda_-})$ where $\lambda_\pm$'s
are the eigenvalues of the matrix  $\Theta^{\dagger} \Theta$ for
general $\Theta$.

\sxn{Non-supersymmetric $SU(5)$ model}

In this section we evaluate the effective potential in the
non-supersymmetric $SU(5)$ gauge theory.  It consists of the
gauge fields $A_M$, the Higgs field $H$ in the fundamental ($\bf 5$)
representation, and fermion multiplets.  We suppose that the gauge fields
and Higgs field live in the bulk five-dimensional spacetime.  Quarks and
leptons are supposed to be confined on the boundary at $y=0$.
If there are additional fermions living in the bulk, they also contribute
to the effective potential.  We include their contributions here for 
generality.
It is known that anomalies may arise at the boundaries
in a five-dimensional model with chiral fermions.\cite{anomaly}
Those anomalies must be cancelled in the four-dimensional effective theory,
for instance,  by  counter terms such as the
Chern-Simons term.\cite{anomaly,KKL}
We assume that the four-dimensional effective theory be anomaly
free.

To be more specific, we adopt
\beqn
&&\hskip -1cm
P_0 =  \mbox{diag}(-1,-1,-1,1,1) ~~,~~
P_1 =  \mbox{diag}(-1,-1,-1,1,1) ~~, \cr
&&\hskip -1cm
U =  \mbox{diag}(1,1,1,1,1) ~~,
\label{case3}
\eeqn
and $+$ sign in (\ref{OrbiBC3}) and (\ref{OrbiBC4}) for the Higgs field
and fermion fields in the bulk.  With the boundary conditions
(\ref{case3}) each field component is  periodic  on $S^1$, and
is either even or odd under reflection $y \go -y$.  Accordingly
a field $\Phi(x,y)$, with $\beta_\Phi =0$ in (\ref{OrbiBC3}), can be
expanded as
\beqn
&& \hskip -1cm
\Phi^+(x,y) = {1\over \sqrt{\pi R}} ~ \phi^+_0(x) +
  \sqrt{{2\over  \pi R} } \sum_{n=1}^{\infty} \phi^+_n (x)
       \cos \bigg( {ny\over R} \bigg)   ~~, \cr
\noalign{\kern 10pt}
&&\hskip -1cm
\Phi^- (x,y) = \sqrt{ {2\over \pi R}}
   \sum_{n=1}^{\infty} \phi_n^- (x) \sin \bigg( {ny\over R}\bigg) ~~,
\label{expansion2}
\eeqn
depending on the parity  of $\Phi$.

To evaluate $V_\eff [A^0]$ for
\beeq
A^0_y = {1 \over 2 g R} \pmatrix{
0 & \Theta \cr \Theta^{\dagger} & 0\cr} ~~,~~
\Theta = \pmatrix{ a & 0 \cr  0 & b \cr 0 & 0\cr} ~~,
\label{Ay2}
\eneq
we need to evaluate $\Tr \ln D_M(A^0) D^M(A^0)$ for various fields.
For a field $B$ in the adjoint representation, for instance,  the
operator  $D_M(A^0) D^M(A^0)$ is given by the relations
\beqn
&&\hskip -1cm
D_M(A^0) D^M(A^0) = \dd_\mu \dd^\mu
     - D_y (A^0_y) D_y (A^0_y)  \cr
\noalign{\kern 10pt}
&&\hskip -1cm
\Tr  B D_y (A^0_y) D_y (A^0_y)   B
= - \Tr  (\dd_y B + ig [A^0_y, B])^2  ~~.
\label{Doperator}
\eeqn
Eigenvalues of $D_y (A^0_y) D_y (A^0_y)$ are found by expanding
$B(x,y)$ in an appropriate basis consisting
of mixture of even and odd functions given by (\ref{expansion2}).

Technical details of computations, with more general boundary
conditions, are given in  Appendices A - D. Gauge fields and ghost
fields have 24 $SU(5)$ components. The spectrum of the eigenvalues
appears in a pair such that it is symmetric under $(a,b) \go (-a,-b)$
as a whole set. As a consequence the sum over $n (\ge 1)$ modes and a
zero mode ($n=0$) is summarized as a sum over $n$ from $-\infty$ to
$+\infty$ for each pair.  The resultant
$V_\eff^{g+gh} [A^0] = V^{g+gh}_\eff (a,b)$ is
\beqn
&&\hskip -1cm
  V_\eff^{g+gh} (a,b)
= - 3 ~ {i \over 2} \int {d^4 p \over (2 \pi)^4}
  ~{1 \over 2 \pi R} ~ \Bigg\{
2 \sum_{n = -\infty}^{\infty} \ln
    \Bigg[ - p^2 +  \bigg( {n \over R} \bigg)^2 \Bigg] \cr
\noalign{\kern 10pt}
&&  + 2 \sum_{n = -\infty}^{\infty} \ln
    \Bigg[ - p^2 +  \bigg( {n - \onehalf a \over R} \bigg)^2 \Bigg]
  + 2 \sum_{n = -\infty}^{\infty} \ln
    \Bigg[ - p^2 +  \bigg( {n - \onehalf b \over R} \bigg)^2 \Bigg] \cr
\noalign{\kern 10pt}
&&
  + 2 \sum_{n = -\infty}^{\infty} \ln
    \Bigg[ - p^2 +  \bigg( {n - \onehalf (a+b) \over R} \bigg)^2 \Bigg]
  + 2 \sum_{n = -\infty}^{\infty} \ln
    \Bigg[ - p^2 +  \bigg( {n - \onehalf (a-b) \over R} \bigg)^2 \Bigg] \cr
\noalign{\kern 10pt}
&& +  \sum_{n = -\infty}^{\infty} \ln
    \Bigg[ - p^2 +  \bigg( {n -  a \over R} \bigg)^2 \Bigg]
  + \sum_{n = -\infty}^{\infty} \ln
  \Bigg[ - p^2 +  \bigg( {n -  b \over R} \bigg)^2 \Bigg]  ~ \Bigg\} ~~.
\label{Veff5}
\eeqn
After Wick rotating the momentum variables, we make use of the formula
\beqn
&&\hskip -1cm
\int {d^{D-1} p_E \over (2 \pi)^{D-1}} {1 \over 2 \pi R}
\sum_{n = -\infty}^{\infty} \ln
  \Bigg[ p_E^2 +  \bigg( {n - x \over R} \bigg)^2 \Bigg]  \cr
\noalign{\kern 15pt}
&&\hskip 0cm
  = - {2 \Gamma(\onehalf D) \over (2 \pi R)^D \pi^{D/2}}
~ f_D(2x)
+ (x{\rm -independent~terms})
\label{integration1}
\eeqn
where
\beqn
&&\hskip -1cm
f_D(x) = \sum_{m= 1}^{\infty} {\cos  m \pi x \over m^D}
= f_D(x+2) = f_D(-x)  \cr
\noalign{\kern 10pt}
&&\hskip -1cm
f_D(0) = \zeta_R(D) ~~,~~
f_D(1) = - (1 - 2^{1-D}) ~ \zeta_R(D)  ~~.
\label{function1}
\eeqn
Here $\zeta_R(z)$ is the Riemann's zeta function.
The effective potential $V_\eff^{g+gh} (a,b)$ becomes
\beqn
&&\hskip -1cm
V_\eff^{g+gh} (a,b)
= -3 C   \bigg\{  f_5(a) + f_5(b)
   +  f_5(a+b) + f_5(a-b) + {1 \over 2}  f_5(2 a)
    + {1 \over 2} f_5(2 b)  \bigg\}  ~ , \cr
\noalign{\kern 10pt}
&&\hskip 1cm
  C = {3 \over 64 \pi^7 R^5} ~~.
\label{Veff6}
\eeqn
Terms independent of $a$ and $b$ have been dropped.

Similarly one can evaluate contributions from the Higgs fields $H$
in {\bf 5} and
fermion fields $\psi$ in {\bf 5} or {\bf 10} in the bulk.
For fermions due care has to be given for the chirality in
four-dimensional  spacetime, which will be detailed in Appendix D.
The result is
\beqn
\noalign{\kern 10pt}
&&\hskip -1cm
\pmatrix{V_\eff^{H_5} (a,b) \cr V_\eff^{\psi_5} (a,b) \cr}
  = \pmatrix{-  C  \cr 2C \cr}
  \Big\{ ~f_5(a - \beta)  + f_5(b - \beta)  ~ \Big\} ~~, \cr
\noalign{\kern 10pt}
&&\hskip -.6cm
V_\eff^{\psi_{10}} (a,b) ~~
= + 2 C  \Big\{ ~f_5( a - \beta) + f_5( b- \beta)   
+ f_5(a +b - \beta) + f_5(a-b - \beta)  ~ \Big\} ~~,
\label{Veff7}
\eeqn
where $\beta$ is  $\beta_\phi$ or $\beta_\psi$ in
(\ref{S1BC1}), and can be either 0 or 1 on $S^1/Z_2$.

\begin{figure}[tbh]
\centering
\leavevmode
\includegraphics[width=8.cm]{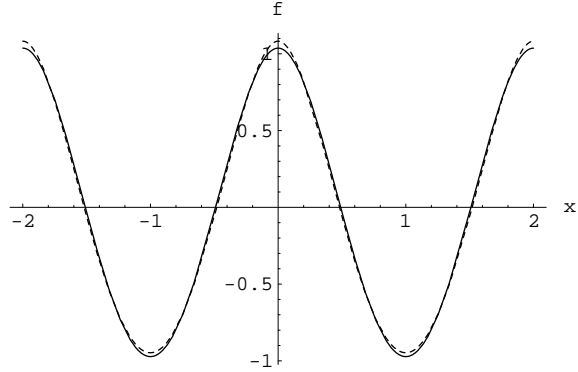}
\caption{$f_5(x)$ (solid line) and $f_4(x)$ (dashed line) in
(\ref{function1}) are plotted. 
$f_D(x)$ depends on $D$ very little.}
\label{fig-f5}
\end{figure}

Suppose that there are  $N_h$ Higgs  fields in {\bf 5},
$N_f^5$  fermions in {\bf 5},  and $N_f^{10}$ fermions {\bf 10}
in the bulk, and also that all the phases $\beta$ vanish.
Then the total effective potential is given by
\beqn
&&\hskip -1cm
V_\eff (a,b) = - C ~
\Bigg\{ N_A \Big[ f_5(a)+ f_5(b) \Big]  \cr
\noalign{\kern 10pt}
&&\hskip 1cm
+N_B ~ \Big[ f_5(a+b) + f_5(a-b) \Big]
+{3 \over 2} \Big[ f_5(2a) + f_5(2b) \Big] ~ \Bigg\}
\label{Veff8}
\eeqn
where
\begin{eqnarray}
&&N_A \equiv 3+N_h-2N_f^5-2N_f^{10} ~~, \cr
&&N_B \equiv 3-2N_f^{10} ~~.
\label{Veff9}
\end{eqnarray}
The effective potential $V_\eff(a,b)$ is symmetric under
$a \leftrightarrow b$, $a \go -a$, and $b \go -b$.   It is periodic
in $a$ and $b$; $V_\eff(a+2,b) = V_\eff(a,b+2) = V_\eff(a,b)$.
If both $N_A$ and $N_B$ are positive, the effective potential is
minimized at $a=b=0$.  At the one loop level the effective
potential is expressed in terms of the function $f_5(x)$ in
(\ref{function1}).  Its behavior is depicted in fig.\ \ref{fig-f5}.

Let us restrict ourselves to the fundamental region
$0 \leq a , b< 2$.  $V_\eff(a,b)$ is stationary  at
$(a,b) = (0,0)$, $(0,1)$, $(1,0)$, $(1,1)$ and $(0.4481, 0.8096)$, 
$(0.8096, 0.4481)$, $(1.5519, 1.1904)$, and $(1.1904,1.5519)$.


Let us compare $V_\eff$ at three distinct points,
$(a,b) = (0,0)$, $(0,1)$,  and $(1,1)$.  $V_\eff(1,0) = V_\eff(0,1) $.
In cases of interest $(0,0)$ and $(1,1)$ correspond to minima, whereas
$(0,1)$ a saddle point.  It follows from (\ref{Veff8}) that
\beqn
&&\hskip -1cm
V_\eff(0,0) = -C \sum_{n=1}^\infty {1\over n^5} ~  [2N_A+2N_B+3] \cr
\noalign{\kern 10pt}
&&\hskip -1cm
V_\eff(0,1) = -C \sum_{n=1}^\infty {1\over n^5} ~
       [(N_A+3) + (-1)^n (N_A+2N_B)] \cr
\noalign{\kern 10pt}
&&\hskip -1cm
V_\eff(1,1) = -C \sum_{n=1}^\infty {1\over n^5} ~
           [(2N_B+3)+ (-1)^n 2N_A ] ~~.
\label{pot1}
\eeqn
and
\beqn
&&\hskip -1cm
V_\eff(0,1) -V_\eff(1,1)  = - 2 C F (N_A-2N_B)   ~~, \cr
\noalign{\kern 10pt}
&&\hskip -1cm
V_\eff(0,0) - V_\eff(0,1) =-2C F (N_A+2N_B) ~~,  \cr
\noalign{\kern 10pt}
&&\hskip -1cm
V_\eff(0,0)-V_\eff(1,1) =- 4 C F N_A  ~~,
\label{pot2}
\end{eqnarray}
where $F  = \sum_{n=1}^\infty (2n-1)^{-5} = (1 - 2^{-5}) \zeta_R(5)$.
Hence, among these three points the  minimum of $V_\eff$
is found at
\beqn
(0,0) &&{\rm if~~}  N_A >0 {\rm ~and~} N_A + 2N_B >0 ~~, \cr
(1,1) &&{\rm if~~}  N_A < 0 {\rm ~and~} N_A - 2 N_B <0 ~~, \cr
(0,1) &&{\rm if~~}  N_A + 2N_B <0 {\rm ~and~}N_A - 2N_B >0 ~~.
\label{minimum1}
\eeqn

\begin{figure}[tbh]
\centering
\leavevmode
\includegraphics[width=8.cm]{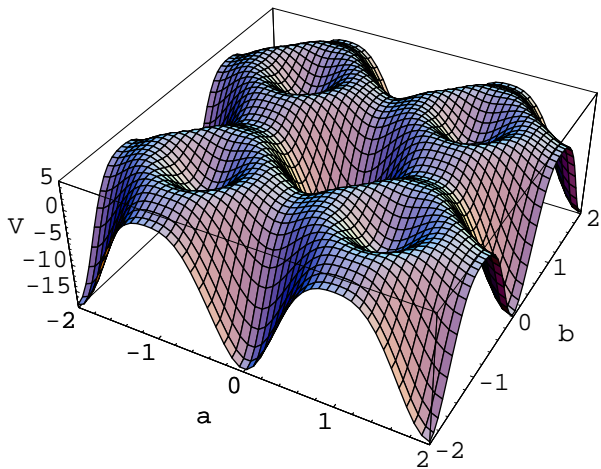}
\caption{$V_\eff(a,b)/C$ in (\ref{Veff8}) for $N_h=1$ and $N_f^5 =
N_f^{10}=0$. The global minimum and local minimum are located at
$(0,0)$ and $(1,1)$, respectively.  The global maxima are located at
$(0.4481, 0.8096)$, $(0.8096, 0.4481)$, $(1.5519, 1.1904)$, and 
$(1.1904,1.5519)$.}
\label{fig-Veff1}
\end{figure}

Let us consider two typical cases.  In the minimal model
$N_h=1$ and $N_f^5 = N_f^{10}=0$.  (There are no fermions living
in the bulk.) Then $N_A = 4$ and $N_B = 3$ so that the global minimum
of $V_\eff$ is found at $(a,b)= (0,0)$.  It is depicted in
fig.\ \ref{fig-Veff1}.

\begin{figure}[tbh]
\centering
\leavevmode
\includegraphics[width=8.cm]{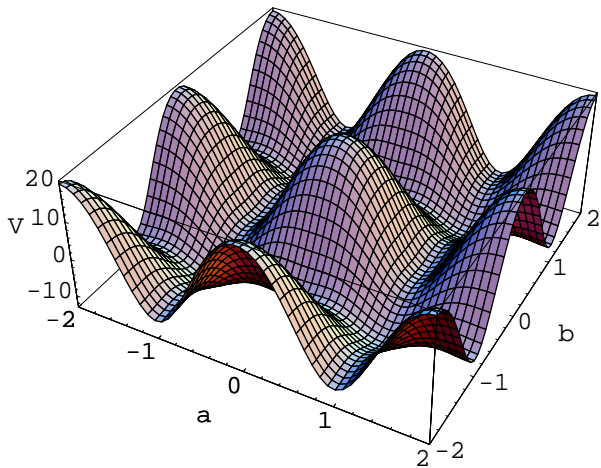}
\caption{$V_\eff(a,b)/C$ in (\ref{Veff8}) for $N_h=1$ and $N_f^5 =
N_f^{10}=3$. The global minimum is located at
$(1,1)$ whereas  $(0,1)$ or $(1,0)$ corresponds to a local minimum.
$(0,0)$ is the global maximum.}
\label{fig-Veff2}
\end{figure}

As another example, consider a case $N_h=1$ and $N_f^5 = N_f^{10}=3$.
Then $N_A = -8$ and $N_B = -3$ so that the minimum is found at
$(a,b) = (1,1)$.
The potential is plotted in fig.\ \ref{fig-Veff2}.  The presence of
bulk fermions drastically changes the symmetry of the theory.

\sxn{Phases in the $SU(5)$ theory}

The symmetry of the true vacuum   for
$(N_h,N_f^5,N_f^{10}) = (1,0,0)$ or $(1,3,3)$ in the model in the
previous section can be found easily.  First consider the case $(1,0,0)$.
The effective potential has the global minimum at $(a,b)=(0,0)$.
The physical symmetry of the theory is the same as the symmetry of the
boundary conditions since $(P_0^\sym, P_1^\sym) = (P_0, P_1)$.  It is
$G_\SM = SU(3) \times SU(2) \times U(1)$.   The extra-dimensional
components of the gauge fields $A_y^a$ ($a=13, \cdots, 24$) acquire
the common mass $M_A$.   Recalling $A_y^{13} \leftrightarrow
a/(gR)$ in (\ref{Ay2}), we find
\beqn
&&\hskip -1cm
M_A^2 = (gR)^2 ~ {\dd^2 V_\eff (a,b) \over \dd a^2} \bigg|_{a=b=0}
= {3g^2 \over 4 \pi^5 R^3} ~ \zeta_R(3) \cr
\noalign{\kern 10pt}
&&\hskip -0.2cm
= {3g_4^2\over  4 \pi^4 R^2}  ~ \zeta_R(3) ~~,
\label{Amass1}
\eeqn
where the four-dimensional coupling constant $g_4^2 = g^2/(\pi R)$ has
been used.  A large mass of $O(g_4/R)$ has been generated by radiative
corrections.  As $\la A_y \ra=0$, the Higgs field and bulk
fermions do not aquire masses by the Hosotani mechanism.  Their masses are
subject to  radiative corrections by other interactions, however.

In the case $(N_h,N_f^5,N_f^{10}) = (1,3,3)$ the global minimum of
$V_\eff$ is located at $(1,1)$.   The Wilson line phases develop
nonvanishing expectation values.  As described in Section 2.5,
the physical symmetry of the theory is given by (\ref{decomposition3}).
$(a,b)=(1,1)$ corresponds to
\beqn
&&\hskip -1cm
\hat w =  2\pi g R  \la A_y \ra  
  = \pi
  \pmatrix{0& 0& 0& 1& 0\cr
           0& 0& 0& 0& 1\cr
           0& 0& 0& 0& 0\cr
           1& 0& 0& 0& 0\cr
           0& 1& 0& 0& 0\cr}  ~~,  \cr
\noalign{\kern 10pt}
&&\hskip -1cm
W = e^{i \hat w}
  = \diag (-1,-1,1,-1,-1) ~~.
\label{Wphase3}
\eeqn
Under a gauge transformation $\Omega(y) = \exp \{ i(y/2\pi R) \hat w\}$,
$\la A_y' \ra=0$ and
\beqn
&&\hskip -1cm
P_0^\sym = P_0 = \diag (-1,-1,-1,1,1) \cr
\noalign{\kern 10pt}
&&\hskip -1cm
P_1^\sym = W P_1 = \diag  (1,1,-1,-1,-1) ~~.
\label{symmetry1}
\eeqn
The physical symmetry is given by $[SU(2)]^2 \times [U(1)]^2$. The color
$SU(3)$ is broken down to $SU(2) \times U(1)$.

At this stage it is appropriate to examine the equivalence classes of
boundary conditions in the $SU(5)$ model.  The notion of the equivalence
class was introduced in Section 2.4, and was claimed in Section 2.5 that
theories belonging to the same equivalence class have the same physics
as a result of the Hosotani mechanism.  We can confirm it in the model
under discussions.

Let us consider several boundary conditions belonging to the same
equivalence class.  Examples are generated by the equivalence relation
(\ref{equiv2}) in an $SU(2)$ subspace.  The transformation
encountered in (\ref{Wphase3}) and (\ref{symmetry1}) also belong to
this category.  The examples we consider are
\beqn
&(\BC 1):&
P_0 =  \diag (-1,-1,-1,1,1) ~~,~~
U =  \diag (1,1,1,1,1) ~~, \cr
&&P_1 =  \diag (-1,-1,-1,1,1) ~~, \cr
&&\hskip 2cm  G_\BC^{(1)} = SU(3) \times SU(2) \times U(1) \cr
\noalign{\kern 10pt}
&(\BC 2):&
P_0 =  \diag (-1,-1,-1,1,1) ~~,~~
U =  \diag (1,-1,1,1,-1) ~~, \cr
&&P_1 =  \diag (-1,1,-1,1,-1) ~~, \cr
&&\hskip 2cm G_\BC^{(2)} = SU(2) \times U(1) \times U(1) \times U(1) \cr
\noalign{\kern 10pt}
&(\BC 3):&
P_0 =  \diag (-1,-1,-1,1,1) ~~,~~
U =  \diag (-1,-1,1,-1,-1) ~~, \cr
&&P_1 =  \diag (1,1,-1,-1,-1) ~~, \cr
&&\hskip 2cm G_\BC^{(3)} = SU(2) \times SU(2) \times U(1) \times U(1) \cr
\noalign{\kern 10pt}
&(\BC 4):&
P_0 =  \diag (-1,-1,-1,1,1) ~~,~~
U =  P_1 P_0  ~~, \cr
\noalign{\kern 10pt}
&&P_1 =  \pmatrix{ - \cos \pi p & 0& 0& -i \sin \pi p& 0\cr
                   0& - \cos \pi q & 0& 0& -i \sin \pi q\cr
                   0& 0& -1 & 0 & 0\cr
                  i \sin \pi p& 0& 0 &    \cos \pi p & 0\cr
                    0& i \sin \pi q & 0& 0& \cos \pi q\cr}
  ~~, \cr
\noalign{\kern 10pt}
&&\hskip 2cm G_\BC^{(4)} = U(1) \times U(1) \times U(1)
\label{boundary1}
\eeqn
$G_\BC$ denotes the symmetry of the boundary conditions at low energies
specified by (\ref{residual2}).  (BC1) is  Case 2, (\ref{case3}).
(BC2) and (BC3) are  special cases of (BC4) with $(p,q)=(0,1)$ and
$(1,1)$, respectively.   (BC3) has been encountered in (\ref{symmetry1}).

All of the boundary conditions listed in (\ref{boundary1})  belong to the
same equivalence class so that the corresponding theories have the
same physics.  The physical symmetry, $H^\sym$ defined in
(\ref{decomposition3}), is determined by the matter content.
$H^\sym = G_\BC^{(1)} = G_\SM$ for $(N_h,N_f^5,N_f^{10}) = (1,0,0)$,
whereas $H^\sym = G_\BC^{(3)}$ for $(N_h,N_f^5,N_f^{10}) =(1,3,3)$.
Although $G_\BC$'s are different, the theories yield the same physical
symmetry by the Hosotani mechanism.

In the rest of this section we shall give detailed accounts of  how the
cases (BC2), (BC3),  and (BC4) lead to the same physics by the dynamics of
the Wilson line  phases.  First we recall that for the boundary conditions
(BC1) the effective potential is given by (\ref{Veff9});
\beqn
&&\hskip -1cm
V_\eff^{\rm BC1} (a,b) = - C ~
\Bigg\{ N_A \Big[ f_5(a)+ f_5(b) \Big]  \cr
\noalign{\kern 10pt}
&&\hskip 1cm
+N_B ~ \Big[ f_5(a+b) + f_5(a-b) \Big]
+{3 \over 2} \Big[ f_5(2a) + f_5(2b) \Big] ~ \Bigg\} ~~.
\label{Veff-BC1}
\eeqn
In the case (BC2) there are six degrees of freedom for the  Wilson
line phases.  They are  given by
\beeq
A^0_y = {1 \over 2 g R}
\pmatrix{0 & \Theta \cr \Theta^{\dagger} & 0\cr}  ~~,~~
\Theta = \pmatrix{ \alpha & 0 \cr  0 & \beta \cr \gamma & 0\cr}
\label{Theta3}
\eneq
where $\alpha, \beta, \gamma$ are complex.
With the aid of the residual symmetry $G_{BC}^{(2)}$ one can take,
without loss of generality, $\alpha=a$, $\beta=b$, and $\gamma=0$
($a,b$: real) for the evaluation of $V_\eff$.  The resultant $A_y^0$
is the same as (\ref{Ay2}).
As described in Appendices C
and D, the change in the boundary conditions amounts to the shift
$b \go b-1$ in the spectrum.  Hence
\beeq
V_\eff^{\rm BC2} (a,b) = V_\eff^{\rm BC1} (a,b - 1)  ~~.
\label{Veff-BC2}
\eneq

Similarly, in the case (BC3) there remain eight degrees of freedom for the
Wilson line phases. They are given by
\beeq
A^0_y = {1 \over 2 g R} \pmatrix{
         0 && \Theta \cr &0\cr
        \Theta^{\dagger} && 0 \cr} ~~,~~
\Theta = \pmatrix{ \alpha & \delta \cr \gamma& \beta \cr}
\label{Theta3}
\eneq
where $\alpha \sim \delta$ are complex.
Again with the aid of  $G_{BC}^{(3)}$ one can take $\alpha=a$, $\beta=b$,
$\gamma=\delta=0$ ($a,b$: real).   The resultant $A_y^0$ is the same as
before.  In this case the spectrum shifts, from the (BC1) case,  as
$a \go a -1$ and $b \go b-1$ so that
\beeq
V_\eff^{\rm BC3} (a,b) = V_\eff^{\rm BC1} (a-1,b - 1)  ~~.
\label{Veff-BC3}
\eneq

It is easy to generalize the argument. With the boundary condition
(BC4) with generic $(p,q)$, there remain only two  degrees of freedom for
Wilson line phases;
\beeq
A^0_y = {1 \over 2 g R} \pmatrix{
         0 &&& a  \cr
         &0 &&&b\cr
         &&0 \cr
         a &&&0 \cr
         &b &&& 0 \cr}
\label{Theta4}
\eneq
where $a, b$ are real.  The computation of the effective potential is 
involved.  The detailed accounts are given in appendices A, C and D.
It turns out that the shift in the spectrum is summarized  by the
replacement 
$a \go a-p$ and $b \go b -q$.   Consequently
\beeq
V_\eff^{\rm BC4} (a,b) \equiv V_\eff^{(p,q)} (a,b)
  = V_\eff^{\rm BC1} (a-p,b - q)  ~~.
\label{Veff-BC4}
\eneq
This  establishes the relation (\ref{Veff3}) in the model under
consideration.

Now one can find the physical symmetry. The global minimum of
$V_\eff^{(p,q)} (a,b)$ is located at $(a,b) = (p,q)$ or $(p-1, q-1)$
($\mbox{mod} ~ 1$) for  $(N_h,N_f^5,N_f^{10})=(1,0,0)$ or $(1,3,3)$,
respectively. In the former case a gauge transformation
\beeq
\Omega(y; p, q) \equiv
\exp \bigg\{ i \, {y\over 2R}\, (p \lambda_{13} + q \lambda_{19} ) \bigg\}
\label{gauge5}
\eneq
brings $\la A_y \ra$ to $\la A_y' \ra = 0$.  $P_0$ remains invariant,
whereas $P_1$ is transformed back to $P_1$ in (BC1).  In other words,
\beqn
&&\hskip -1cm
{\rm for ~} (N_h, N_f^5,N_f^{10})=(1,0,0) \cr
\noalign{\kern 10pt}
&&\hskip -.5cm
P_0^\sym =  \diag (-1,-1,-1, 1,1) ~~, \cr
&&\hskip -.5cm
P_1^\sym =  \diag (-1,-1,-1,1,1) ~~, \cr
&&\hskip -.5cm
U^\sym  = \diag (1,1,1,1,1) ~~, \cr
\noalign{\kern 10pt}
&&\hskip -.5cm
H^\sym   = SU(3) \times SU(2) \times U(1)
\label{symmetry2}
\eeqn
irrespective of the values of $(p,q)$.
Similarly, for  $(N_h,N_f^5,N_f^{10})=(1,3,3)$ a gauge transformation
$\Omega(y; p-1, q-1)$ brings $\la A_y \ra$ to $\la A_y' \ra = 0$.
The resulting symmetry is
\beqn
&&\hskip -1cm
{\rm for ~} (N_h, N_f^5,N_f^{10})=(1,3,3) \cr
\noalign{\kern 10pt}
&&\hskip -.5cm
P_0^\sym =  \diag (-1,-1,-1,1,1) ~~, \cr
&&\hskip -.5cm
P_1^\sym =  \diag (1,1,-1,-1,-1) ~~, \cr
&&\hskip -.5cm
U^\sym = \diag (-1,-1,1,-1,-1) ~~, \cr
\noalign{\kern 10pt}
&&\hskip -.5cm
H^\sym   = SU(2) \times SU(2) \times U(1) \times U(1)
\label{symmetry3}
\eeqn
independent of $(p,q)$.

As far as two theories belong to the same  equivalence class of boundary
conditions,   the physical symmetry is the same.  It is guaranteed by the
Hosotani mechanism as stated in Section 2.  The symmetry depends on the
matter content in the theory.  Dynamical rearrangement of gauge symmetry
has taken place.   We summarize the result in Table \ref{table-sym}.
We remark that the number of Wilson line phases depends on the boundary
conditions chosen.  This does not mean, however, that the total number of
degrees of freedom in the theory varies with the boundary conditions.
Wilson line phases are zero modes ($y$-independent modes) in $A_y$'s.  As
explained in Appendix A, some components of $A_y$'s have mode expansion
in $\big\{ \cos [(n+ p)y/R] \big\}$ or $\big\{ \cos [(n+ \onehalf  p)y/R]
\big\}$ when the boundary conditions are given by (BC4) in
(\ref{boundary1}). There appear zero modes when $n+p$ or $n+ \onehalf  p$
can be zero for  an integral $n$.  The number of degrees of freedom is
unchanged as the value of $p$ changes.

\vskip .5cm
\begin{table}[hb]
\begin{center}
\begin{tabular}{|c|c|c|}     \hline
\multicolumn{1}{|c|} {matter content} &{minimum of $V_\eff$}
        & {physical symmetry}  \\
\multicolumn{1}{|c|} {$(N_h,N_f^5,N_f^{10})$} & {$(~a~,~b~)$}
        &  {$H^\sym$} \\ \hline
\multicolumn{1}{|c|} {$(~1~,~0~,~0)$} & {$(~p~,~q~)$}  &
{$SU(3)\times SU(2)\times U(1)$} \\ \hline
\multicolumn{1}{|c|} {$(~1~,~3~,~3~)$} & {$(p-1, q-1)$}  &
{$SU(2)\times SU(2)\times U(1)\times U(1)$} \\  \hline
\end{tabular}
\end{center}
\caption[table-sym]{Physical symmetry is summarized in the
non-supersymmetric $SU(5)$ theory with the boundary conditions (BC4) in
(\ref{boundary1}).  The physical symmetry is determined by the matter
content, but is independent of the parameters $(p,q)$ in the
boundary conditions.}
\label{table-sym}%
\end{table}

\vskip .5cm

\sxn{Supersymmetric $SU(5)$ model}

\ignore{In the non-supersymmetric $SU(5)$ gauge theory discussed in the 
previous
sections the triplet-doublet mass splitting problem for the Higgs
field in the fundamental representation is solved at the tree level
by the orbifold boundary condition (\ref{case3}).  However,
radiative corrections generally destroy the relation; both
triplet and doublet components would acquire large mass corrections,
unless fine-tuning of parameters were done.
One  way to preserve the relation beyond the tree level is
to resort to supersymmetry (SUSY) as Kawamura has done in his
original  $SU(5)$ gauge models on $M^4 \times
S^1/Z_2$.\cite{YK2,YK3,Hall1}.   We shall reexamine this problem in the SUSY
$SU(5)$ model with the orbifold boundary condition (\ref{case3}).}

In the non-supersymmetric $SU(5)$ gauge theory discussed in the previous
sections, the triplet-doublet mass splitting for the Higgs
field in the fundamental representation is realized at the tree level
by the orbifold boundary condition (\ref{case3}).
However, we have not taken into account a contribution from the
potential $V(\phi,  \psi)$ at the tree level.
In general, there is a mass term $m_{\phi}^2 |\phi|^2$, which is
subject to   radiative corrections.
It is natural to suppose that the magnitude of $m_{\phi}$ is as big
as the  unification  scale in
grand unified theories and   can become as large as a cutoff scale by
radiative  corrections due to inherent quadratic divergences.
In these circumstances both triplet and doublet components would acquire
large mass  corrections,
unless fine-tuning of parameters were exercised.
One way to preserve the large mass splitting between triplet and
doublet Higgs fields
at and beyond the tree level is to resort to supersymmetry (SUSY) as 
Kawamura has proposed in his
original $SU(5)$ gauge models on 
$M^4 \times (S^1/Z_2)$.\cite{YK2, Hall1}.
The mass term of Higgs multiplets is forbidden by $U(1)_R$ symmetry at the 
tree level
and there is no quadratic divergence to alter the magnitude of scalar 
masses thanks to SUSY.
Bearing these advantages in mind,
we shall investigate features of the SUSY
$SU(5)$ model with the orbifold boundary condition (\ref{case3}).

One general comment is in order.  If the boundary conditions
(\ref{case3}) is adopted, there appear Wilson line degrees of freedom
as in the non-supersymmetric theory.  There are dynamics of
those Wilson line phases.  However, if supersymmetry remains
exact and unbroken, the effective potential $V_\eff$ for the Wilson line
phases remains flat at the one loop level, i.e.\ there remain degenerate
vacua.

To have nontrivial dynamics supersymmetry must be broken, either
spontaneously or by soft breaking terms.   On multiply connected
manifolds there is a natural way of introducing soft SUSY breaking.
Scherk and Schwarz noted that distinct twisting along,
say, $S^1$, for bosons and fermions can be implemented without spoiling
good properties of  SUSY theories.\cite{SS}  This Scherk-Schwarz
mechanism can be exploited in theories on 
  $M^4 \times (S^1/Z_2)$.\cite{SS3,Barbieri1}  Takenaga has examined
the Hosotani mechanism in supersymmetric gauge theories on  $M^3 \times
S^1$ with Scherk-Schwarz SUSY breaking.\cite{Takenaga}  Gersdorff and 
Quiros have
shown that the Scherk-Schwarz SUSY breaking can be realized as the Hosotani
mechanism in the gauged supergravity model as well.\cite{Quiros1}
We shall adopt the Scherk-Schwarz mechanism for the SUSY breaking,
which makes the evaluation of the effective potential easy.

We start to specify the content of the SUSY $SU(5)$ model on
$M^4 \times (S^1/Z_2)$.  We take, as an example, the model investigated
in ref.\ \cite{Barbieri1} modified in the orbifold boundary conditions.
$N=1$ SUSY in five-dimensional space-time corresponds to $N=2$ SUSY in 
four-dimensional
ones.  A five-dimensional gauge multiplet
\beeq
{\cal{V}}=(A^M, \lambda, \lambda', \sigma).
\label{multiplet1}
\eneq
is decomposed, in four dimensions,  to a vector super-field
and a chiral super-field
\beeq
V=(A^\mu, \lambda) ~~,~~ \Sigma = (\sigma + iA^y, \lambda') ~~.
\label{multiplet2}
\eneq
We introduce  hypermultiplets in  fundamental representation
({\bf 5}),
\beeq
{\cal H}=(h, h^c{}^\dagger, \tilde{h},
   \tilde{h}^c{}^\dagger) ~~,~~
\overline{{\cal H}}=(\overline{h}, \overline{h}^c{}^\dagger,
   \tilde{\overline{h}}, \tilde{\overline{h}}^c{}^\dagger) ~~,
\label{multiplet3}
\eneq
which  are decomposed into chiral superfields as
\beqn
&&\hskip -1cm
H=(h, \tilde{h}) ~~,~~
\overline{H}=(\overline{h}, \tilde{\overline{h}})  \cr
&&\hskip -1cm
H^c=(h^c, \tilde{h}^c)  ~~,~~
  \overline{H^c}=(\overline{h}^c, \tilde{\overline{h}^c})
\label{multiplet4}
\eeqn
where $H$ ($\overline{H}$)and $H^c$ ($\overline{H^c}$)
  have conjugated transformation  under the gauge group $SU(5)$.

Next we write down boundary conditions for each field based on the boundary
conditions (\ref{case3}).
Under the $Z_2$ reflection at $y=0$, each superfield transforms such that
\beqn
&&\hskip -1cm
\pmatrix{ V \cr \Sigma\cr} (x, -y)
= P_0 ~\pmatrix{ V \cr -\Sigma\cr}  (x, y) ~ P_0^\dagger \cr
&&\hskip -1cm
\left(
\begin{array}{cc}
  H  & \overline{H}  \\
  H^c{}^\dagger  & \overline{H}^c{}^\dagger
\end{array}
\right) (x, -y)  =
  P_0  \left(
\begin{array}{cc}
H  & -\overline{H} \\
  -H^c{}^\dagger  & \overline{H}^c{}^\dagger
\end{array}
\right) (x,y)  ~~ ,
\label{SBC1}
\end{eqnarray}
where we take the opposite parity between
  ${\mathcal H}$ and $\overline{\mathcal H}$.
In this case the supersymmetric Higgs mass term called the $\mu$-term
can  be derived by  twisting boundary conditions on
$S^1$.\cite{Barbieri1}
If the $\mu$ term is induced
  by another mechanism such as the Giudice-Masiero  mechanism,\cite{Giudice}
  there is no need to assign opposite parity between
  ${\mathcal H}$ and $\overline{\mathcal H}$.

\ignore{
Here $P_0$ is a $5 \times 5$ matrix acting in the $SU(5)$ space.
If $P_0=I$, then  the $\Sigma$ and $H^c$  do not have  zero
modes,  and there remain no degrees of freedom of  Wilson line phases.
As in the nonsupersymmetric models discussed in the previous sections we
adopt  non-trivial $P_0$ in (\ref{case3}). }

For the shift by $2 \pi R$ on $S^1$,
we impose the following boundary condition on each field a la Scherk and 
Schwarz,\cite{SS}
\begin{eqnarray}
&&\hskip -1cm
A^M(x^\mu, y+2 \pi R)= U ~ A^M(x^\mu, y) ~ U^\dagger ~~, \cr
\noalign{\kern 10pt}
&&\hskip -1cm
\left(
\begin{array}{c}
\lambda \\
\lambda'
\end{array}
\right) (x, y+2\pi R)
= e^{-2\pi i \beta \sigma_2} ~
U ~\left(
\begin{array}{c}
\lambda \\
\lambda'
\end{array}
\right) (x,y) ~U^\dagger ~~,  \cr
\noalign{\kern 10pt}
&&\hskip -1cm
\sigma(x,y+2\pi R)= U~ \sigma(x,y) ~ U^\dagger ~~,\cr
\noalign{\kern 10pt}
&&\hskip -1cm
\left(
\begin{array}{cc}
  h  & \overline{h}  \\
  h^c{}^\dagger  &
              \overline{h}^c{}^\dagger
\end{array}
\right) (x, y+2\pi R)
= e^{-2\pi i \beta \sigma_2} ~ U ~
\left(
\begin{array}{cc}
  h  & \overline{h}  \\
  h^c{}^\dagger  &
              \overline{h}^c{}^\dagger
\end{array}
\right) (x,y) ~~,  \cr
\noalign{\kern 10pt}
&&\hskip -1cm
\left(
\begin{array}{cc}
  \tilde{h}  & \tilde{\overline{h}}
\\
  \tilde{h}^c{}^\dagger  &
              \tilde{\overline{h}^c}{}^\dagger
\end{array}
\right) (x, y+2\pi R) =
U ~ \left(
\begin{array}{cc}
  \tilde{h}  & \tilde{\overline{h}}  \\
  \tilde{h}^c{}^\dagger  &
              \tilde{\overline{h}^c}{}^\dagger
\end{array}
\right) (x,y) ~~,
\label{SBC2}
\end{eqnarray}
where $\sigma_2$ is  the Pauli  matrix in $SU(2)_R$.  $\beta$ is real.
The five-dimensional action possesses $SU(2)_R$ symmetry.
With a nonvanishing $\beta$, there appear soft SUSY breaking mass terms
for gauginos and scalar fields in four-dimensional theory as will be seen 
below.

Boundary conditions under $Z_2$ reflection at $y=\pi R$ follow from
(\ref{SBC1}) and (\ref{SBC2}) by use of generic arguments in 2.1.,
\begin{eqnarray}
&&\hskip -1cm
A^\mu(x^\mu, \pi R-y)= P_1 ~ A^\mu(x^\mu, \pi R+y) ~ P_1^\dagger ~~, \cr
\noalign{\kern 10pt}
&&\hskip -1cm
A^y(x^\mu, \pi R-y)= -P_1 ~ A^y(x^\mu, \pi R+y) ~ P_1^\dagger ~~, \cr
\noalign{\kern 10pt}
&&\hskip -1cm
\left(
\begin{array}{c}
\lambda \\
\lambda'
\end{array}
\right) (x, \pi R-y)
= e^{-2\pi i \beta \sigma_2} ~
P_1 ~\left(
\begin{array}{c}
\lambda \\
-\lambda'
\end{array}
\right) (x, \pi R+y) ~P_1^\dagger ~~,  \cr
\noalign{\kern 10pt}
&&\hskip -1cm
\sigma(x,\pi R-y)= -P_1~ \sigma(x,\pi R+y) ~ P_1^\dagger ~~,\cr
\noalign{\kern 10pt}
&&\hskip -1cm
\left(
\begin{array}{cc}
  h  & \overline{h}  \\
  h^c{}^\dagger  &
              \overline{h}^c{}^\dagger
\end{array}
\right) (x, \pi R-y)
= e^{-2\pi i \beta \sigma_2} ~ P_1 ~
\left(
\begin{array}{cc}
   h  & -\overline{h}  \\
  -h^c{}^\dagger  &
              \overline{h}^c{}^\dagger
\end{array}
\right) (x,\pi R+y) ~~,  \cr
\noalign{\kern 10pt}
&&\hskip -1cm
\left(
\begin{array}{cc}
  \tilde{h}  & \tilde{\overline{h}}
\\
  \tilde{h}^c{}^\dagger  &
              \tilde{\overline{h}^c}{}^\dagger
\end{array}
\right) (x, \pi R-y) =
P_1 ~ \left(
\begin{array}{cc}
   \tilde{h}  & -\tilde{\overline{h}}  \\
  -\tilde{h}^c{}^\dagger  &
              \tilde{\overline{h}^c}{}^\dagger
\end{array}
\right) (x,\pi R+y) ~~,
\label{SBC3}
\end{eqnarray}
where $P_1 = P_1^\dagger = UP_0^\dagger = U P_0$.

From the above boundary conditions (\ref{SBC1}), (\ref{SBC2}) and 
(\ref{SBC3}), mode expansions of each field are obtained.
We adopt the boundary condition (\ref{case3}), or (BC1) in (\ref{boundary1}). 
The components
\beqn
&&\hskip -1cm
A_\mu^{a} \quad (a = 1 \sim 12) ~~,~ ~ 
A_y^{b} ~,~ \sigma^b \quad  (b = 13 \sim 24) ~, \cr
&&\hskip -1cm 
\widetilde{h}_i ~,~ 
\widetilde{\overline{h}^c_i}{}^\dagger \quad (i = 4,5) ~,
\quad
\widetilde{\overline{h}^c_j} ~,~ 
\widetilde{h}^c_j{}^\dagger \quad (j = 1,2,3)
\label{SUSYcos}
\eeqn
have the same expansion as $\Phi^+(x,y)$ in (\ref{expansion2}), whereas
the components
\beqn
&&\hskip -1cm 
A_\mu^{b} \quad (b = 13 \sim 24)~, \quad 
A_y^{a} ~, \sigma^a  \quad  (a = 1 \sim 12) \cr
&&\hskip -1cm 
\widetilde{h}_j ~,~ \widetilde{\overline{h}^c_j}{}^\dagger \quad (j = 1,2,3)
~, \quad
\widetilde{\overline{h}^c_i} ~,~ \tilde{h}^c_i{}^\dagger  \quad(i = 4,5) 
\label{sin}
\eeqn
have the same expansion as $\Phi^-(x,y)$ in (\ref{expansion2}).
The gaugino and scalar fields have twist in the $SU(2)_R$ space.
Their mode expansions are of the type discussed in Appendix A.
Gauginos are expanded as 
\beqn
&&\hskip -1cm
\pmatrix{ \lambda^a (x, y)\cr {\lambda'}^a (x, y) \cr}
 = {1\over \sqrt{\pi R}} \sum_{n =-\infty}^{\infty} \lambda^a_n (x) 
\pmatrix{ \mybig \cos {(n + \beta)y\over R} \cr 
          \mybig \sin {(n + \beta)y\over R} \cr} ~~, \cr
\noalign{\kern 10pt}
&&\hskip -1cm
\pmatrix{ \lambda^b (x, y)\cr {\lambda'}^b (x, y) \cr}
 = {1\over \sqrt{\pi R}} \sum_{n =-\infty}^{\infty} \lambda^b_n (x) 
\pmatrix{ \mybig \sin {(n-\beta)y\over R} \cr 
          \mybig \cos {(n-\beta)y\over R} \cr} ~~, 
\label{SUSYexpansion1}
\eeqn
for $a= 1 \sim 12$ and $b = 13 \sim 24$,  whereas Higgs fields are 
expanded as 
\beqn
&&\hskip -1cm
\pmatrix{  h_j  & \overline{h}_j  \cr
           h^c_j{}^\dagger  &  \overline{h}^c{}^\dagger_j \cr} (x, y)
 = {1\over \sqrt{\pi R}} \sum_{n =-\infty}^{\infty} 
\pmatrix{
h_{j,n}(x) \mybig \sin {(n-\beta)y\over R} 
    & \overline{h}_{j,n} (x) \mybig \cos {(n+\beta)y\over R} \cr
h_{j,n}(x) \mybig \cos {(n-\beta)y\over R} 
    & \overline{h}_{j,n} (x) \mybig \sin {(n+\beta)y\over R} \cr} \cr
\noalign{\kern 20pt}
&&\hskip -1cm
\pmatrix{  h_i  & \overline{h}_i  \cr
           h^c_i{}^\dagger  &  \overline{h}^c{}^\dagger_i \cr} (x, y)
 = {1\over \sqrt{\pi R}} \sum_{n =-\infty}^{\infty} 
\pmatrix{
h_{i,n}(x) \mybig \cos {(n+\beta)y\over R} 
    & \overline{h}_{i,n} (x) \mybig \sin {(n-\beta)y\over R} \cr
h_{i,n}(x) \mybig \sin {(n+\beta)y\over R} 
    & \overline{h}_{i,n} (x) \mybig \cos {(n-\beta)y\over R} \cr} 
\label{SUSYexpansion2}
\eeqn
for $j = 1,2,3$ and $i = 4,5$.

When the above mode expansions are inserted, 
the following mass terms appear for gauginos and Higgs scalars  upon
compactification from the kinetic terms of the five-dimensional theory;
\begin{eqnarray}
&&\hskip -1cm
{\cal L}_{\rm soft}^{}= {1 \over 2}
\left(\sum_{a = 1}^{12} \sum_{n = -\infty}^{\infty} {n + \beta \over R}
   (\lambda_n^a)^2 \right.
- \sum_{b = 13}^{24} \sum_{n = -\infty}^{\infty}
\left.{n - \beta \over R} (\lambda_n^b)^2 + \mbox{h.c.}\right) \cr
\noalign{\kern 10pt}
&&\hskip 1.cm
- \sum_{j=1}^3 \sum_{n = -\infty}^{\infty}
\left( \left({n - \beta \over R}\right)^2 |{h}_{j,n}(x)|^2 + \right.
\left. \left({n + \beta \over R}\right)^2 |\overline{h}_{j,n}(x)|^2) \right)
\cr
\noalign{\kern 10pt}
&&\hskip 1.cm
- \sum_{i=4}^5 \sum_{n = -\infty}^{\infty}
\left( \left({n + \beta \over R}\right)^2 |{h}_{i,n}(x)|^2 + \right.
\left. \left({n - \beta \over R}\right)^2|\overline{h}_{i,n}(x)|^2) \right)
~~ ,
\label{softmass}
\end{eqnarray}
where $(\lambda_n^a)^2 = (\lambda_n^a)_\alpha^{}
\ep^{\alpha\beta} (\lambda_n^a)_\beta^{}$.  At the same time gauge bosons
and Higgsinos acquire masses, $n/R$,  upon compactification.  Hence the
supersymmetry is broken explicitly by the twisted boundary condition in the
$SU(2)_R$ space on $S^1$.  The zero modes of gauginos and Higgs scalars,
some of which constitute 
 the MSSM, have   non-vanishing masses, $\beta/R$.
Those masses are interpreted as the soft SUSY breaking masses.
If there is an additional $U(1)$ twisting 
in the flavor $SU(2)$ space of  $H_i$ and $\overline{H_i}$, there
appear $\mu$ terms proportional to this $U(1)$ phase.\cite{Barbieri1}
For the sake of simplicity this twisting is suppressed in the present paper.

\begin{figure}[tbh]
\centering  \leavevmode
\includegraphics[width=8.cm]{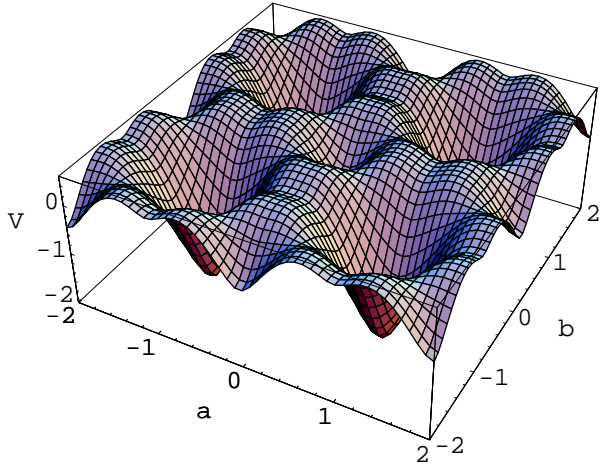}
\caption{$V_\eff(a,b)/2C$ in (\ref{SUSYeffpot}) for $N_h=2$ and $\beta=0.1$
is depicted.
The global minimum is located at
$(1,1)$ whereas  $(0,0)$  corresponds to a local minimum.
For $N_h=1$ there are degenerate global minima at  $(0,0)$ and $(1,1)$.}
\label{fig-Veff-susy}
\end{figure}

With all these mode expansions the effective potential at the one loop level
is evaluated in a similar manner to that in the nonsupersymmetric model.
One finds that %
\beqn
&&\hskip -1.5cm 
V_\eff (a,b)
= -2C \Bigg\{ 2(1- N_h)
[f_5(a)-{1\over2}f_5(a+2\beta)-{1\over2}f_5(a-2\beta) \cr
\noalign{\kern 5pt}
&&\hskip 1.5cm
 +f_5(b)-{1\over2}f_5(b+2\beta)-{1\over2}f_5(b-2\beta)] \cr
\noalign{\kern 5pt}
&&\hskip 1.5cm
+2f_5(a+b)-f_5(a+b+2\beta) -f_5(a+b-2\beta) \cr
\noalign{\kern 5pt}
&&\hskip 1.5cm
 +2f_5(a-b)-f_5(a-b+2\beta)-f_5(a-b-2\beta) \cr
\noalign{\kern 5pt}
&&\hskip 1.5cm
+f_5(2a)-{1\over2}f_5(2(a+\beta))-{1\over2}f_5(2(a-\beta))  \cr
\noalign{\kern 5pt}
&&\hskip 1.5cm
+f_5(2b)-{1\over2}f_5(2(b+\beta))-{1\over2}f_5(2(b-\beta)) \Bigg\} \cr
\noalign{\kern 5pt}
&&\hskip -0.5cm
= - 2 C \sum_{n = 1}^{\infty}
{1 \over n^5} (1-\cos 2\pi n\beta) 
~ \bigg\{ 2(1- N_h)(\cos \pi na +\cos \pi nb) \cr
\noalign{\kern 5pt}
&&\hskip 3.0cm
 +4\cos \pi na  \cos \pi nb   +\cos 2\pi na +\cos 2\pi nb \bigg\} 
\label{SUSYeffpot}
\eeqn
where $N_h$ indicates the number of the set of hyper-multiplets
  ${\mathcal H}+\overline{\mathcal H}$.   Note that $V_\eff(a,b) = 0$ 
for $\beta=0$.  
The global minimum of $V_\eff$  is located  at $(a,b)=(0,0)$ or $(1,1)$
depending on  $N_h$.
The difference in the height of the potential
at these two points is 
\beqn
&&\hskip -1cm 
V_\eff (0,0)-V_\eff (1,1)
= - 16(1- N_h) C \sum_{n = 1}^{\infty}
    {1 \over (2n-1)^5}~ \Big\{ 1 - \cos 2\pi (2n-1) \beta \Big\} \cr
\noalign{\kern 10pt}
&&\hskip 1.0cm
\sim - 32 \pi^2 (1-  N_h) C \beta^2  
  \sum_{n=1}^\infty  {1\over (2n-1)^3}  \qquad \hbox{for small } \beta ~.
\label{V_diff_SUSY}
\eeqn
Irrespective of the value of the SUSY breaking parameter $\beta$ 
 the point $(a,b)=(0,0)$ is
  the global minimum of $V_\eff(a,b)$  when $N_h=0$ and $N_h=1$.
The numerical study shows that
\begin{eqnarray}
\begin{array}{c|c|cc}
N_h     & (a,b)=(0,0) & (a,b)=(1,1)    &    \\ \cline{1-3}
0       & \mbox{global min.}  & \mbox{local min.}  &    \\
1 & \mbox{degenerate global min.}   & \mbox{degenerate global min.} &  \\
2\sim 4 & \mbox{local min.}   & \mbox{global min.} &    \\
5\sim   & \mbox{unstable}   & \mbox{global min.} &    \\
\end{array}
\label{SUSYmin}
\end{eqnarray}
for  small  $|\beta|$.
$\beta$ should be of order $10^{-14}$ on the phenomenological 
ground for the soft SUSY breaking masses to be   $O(1)$TeV in 
Eq.(\ref{softmass}).
In fig.\ \ref{fig-Veff-susy}   $V_\eff(a,b)$ for $N_h=2$ and $N_f^5 =
N_f^{10}=0$ is  depicted.
At the global minimum $(a,b)=(1,1)$  the gauge symmetry is dynamically
broken  to $SU(2)\times SU(2) \times U(1) \times U(1)$.
Color $SU(3)$ is broken, which is unacceptable phenomenologically.
In the theory with $N_h=1$ and $N_f^5 =N_f^{10}=0$, both $(a,b)=(0,0)$ and
$(a,b)=(1,1)$ are global minima of the effective potential.  At 
$(a,b)=(0,0)$ the symmetry is $SU(3)\times SU(2) \times U(1)$.

Physical symmetry is summarized in Table II.  No more than one 
Hyper-multiplet Higgs fields can   live in the bulk in the SUSY
$SU(5)$ model with the boundary conditions (\ref{case3}) 
to maintain $G_\SM$ as physical symmetry.
To have  triplet-doublet splitting, the weak colorless Higgs chiral
supermultiplets can be  located
on the boundary brane not in company with the colored ones, as
 the gauge symmetry of the four-dimensional boundary brane is not 
$SU(5)$ but
$SU(3) \times SU(2) \times U(1)$.
Such an idea for triplet-doublet splitting has been proposed by Hebecker 
and March-Russell\cite{HM}
in the scenario that our world is not $SU(5)$-symmetric  but 
$SU(5)$-violating brane
in the SUSY SU(5) model  equivalent to the one with the boundary 
conditions (\ref{case1}).

\begin{table}[ht]
\begin{center}
\begin{tabular}{|c|c|c|}     \hline
\multicolumn{1}{|c|} {Higgs content} &{minimum of $V_\eff$}
        & {physical symmetry}  \\
\multicolumn{1}{|c|} {$N_h$} & {$(a,b)$}
        &  {$H^\sym$} \\ \hline
\multicolumn{1}{|c|} {$0$} & {$(~0~,~0~)$}  &
{$SU(3)\times SU(2)\times U(1)$} \\ \hline
\multicolumn{1}{|c|} {$1$} & 
      {${\mybig (~0~,~0~)\atop \mbig (~1~, ~1~)}$}  &
     {${\mybig SU(3)\times SU(2)\times U(1) 
        \atop \mbig SU(2)\times SU(2)\times U(1)\times U(1)}$} \\  \hline
\multicolumn{1}{|c|} {$2$} & {$(~1~, ~1~)$}  &
{$SU(2)\times SU(2)\times U(1)\times U(1)$} \\  \hline
\end{tabular}
\end{center}
\caption[table-sym]{Physical symmetry is summarized in the
supersymmetric $SU(5)$ theory with the boundary conditions
(\ref{case3}).  The SM gauge symmetry is preserved only when
$N_h=0$ or $N_h=1$. }
\label{table-sym-SUSY}%
\end{table}

The mass of $A_y^a$ $(a = 13 \sim 24)$ in the $SU(3) \times SU(2) \times 
U(1)$ vacuum  is calculated to be
\beeq
m^2= (gR)^2  {\dd^2 V_\eff (a,b) \over \dd a^2} \bigg|_{a=b=0}
\simeq  {3g_4^2\beta^2 \over 32R^2} ~ (5- N_h)
   ~ \bigg(1-{2\over3} \log 2\pi\beta  \bigg)  ~~,
\label{mass00}
\eneq
for small $\beta$ where use has been made of
\beeq
\sum_{n = 1}^{\infty}
    {\cos(n\beta) \over n^3} \simeq \zeta_R (3)+
     {\beta^2 \over 2}\log{\beta}-{3\over4}\beta^2 ~.
\eneq
We find that the $A_y$ components containing Wilson line degrees of
freedom acquire masses by  quantum corrections and
their magnitude is of order the soft SUSY breaking   $\beta/R$.
In passing, Eq.\ (\ref{mass00}) shows that large $N_h \geq 3$ makes
  the mass square, namely the curvature at origin,  be negative,
which is consistent  with the numerical results in the 
table in Eq.\ (\ref{SUSYmin}).

Before closing this section, we would like to comment on
 proton stability in this model.
We suppose that quarks and leptons are localized  on the  brane at $y=0$.
In the minimal case $N_h = 0$, there are no dangerous processes inducing 
proton decay by dimension 5
operators as there lack  colored Higgs multiplets.
In the case  $N_h = 1$, there may appear colored Higgs multiplets with
masses at the  TeV scale.
Their existence, however,  does not threaten the stability of proton
thanks to  the $U(1)_R$ symmetry on the brane.
Matter fields  have a unit   $U(1)_R$ charge so that 
 dimension 5 operators such as $[QQQL]_F$ and 
$[\bar{U}\bar{U}\bar{D}\bar{E}]_F$ are forbidden.
Dimension 4 operators, which trigger 
rapid proton decay, are also forbidden.
The effective dimension 6 operators induced through the exchange of 
four-dimensional $X$ and $Y$ gauge bosons,
  $[Q^{\dagger} \bar{U}Q^{\dagger} \bar{E}]_D$ and
  $[Q^{\dagger} \bar{U}L^{\dagger} \bar{D}]_D$,
 are also absent because $X$ and $Y$ gauge bosons do not live in the 
four-dimensional hypersurface.
There are no diagrams involving the exchange of scalar fields $X_5^{a}$ and 
$Y_5^{a}$ $(a = 13 \sim 24)$
unless such light mirror particles  as $Q^c$ and $L^c$ exist on the 
four-dimensional brane.
We conclude that the proton life time
is long enough in our model.

We stress that in the SUSY model with $N_h=1$, which is most interesting 
from the phenomenological  viewpoint, the color-conserving vacuum 
is legitimate choice of the nature.

\sxn{False vacuum decay}

In the preceding sections we have found that the effective potential
$V_\eff(a,b)$ is minimized at $(a,b)=(0,0)$ or $(1,1)$, depending on
the matter content.  The true vacuum corresponds to the global
minimum of $V_\eff(a,b)$.  As displayed in figs.\ \ref{fig-Veff1},
\ref{fig-Veff2}, and \ref{fig-Veff-susy} there appears a local
minimum of the potential in each case.  One may wonder what would
happen if the universe were trapped in the local minimum, or in the
false vacuum.

The false vacuum would eventually decay to the true vacuum by  tunneling.
How long does it take before the system decays to settle in the true
vacuum?  The decay rate is estimated in the semiclassical
method.\cite{Coleman}  The transition probability
per unit volume per unit time is estimated as
\begin{eqnarray}
\frac{\Gamma}{V} \sim A e^{-S_E}
\end{eqnarray}
where $S_E$ is the Euclidean action of the bounce solution.
The front coefficient  $A$ is  $\sim
M^{4}$  where $M$ is a typical mass scale of the problem.
We shall consider two cases;  the nonsupersymmetric and
supersymmetric models with
$(N_h,N_f^5,N_f^{10}) = (1,0,0)$.  In the former case $(a,b)=(0,0)$
and $(1,1)$  corresponds to the true and false vacua, respectively.
It is the other way around in the latter case.

To find a bounce solution we first note that the size $R$ of the fifth
dimension is so small that a bounce solution is expected to be
approximately uniform in the fifth dimension.  It is four-dimensional.
The relevant fields are $A_y^{13}$ and $A_y^{19}$.  As seen from
the shape of the potential in fig.\  \ref{fig-Veff1} or
fig.\ \ref{fig-Veff-susy}, the transition from the false vacuum to
the true vacuum takes place mostly along a straight line in
the $a$-$b$ space connecting $(0,0)$ and $(1,1)$.   Accordingly we
restrict ourselves to $A_y^{13} = A_y^{19}$ and introduce
a field $\phi(x)$ by
\beeq
\phi(x) = {1\over 2} \, gR \, (A_y^{13} + A_y^{19}) ~~.
\label{bounce-phi}
\eneq
The bounce solution is a function of
$\rho = R^{-1} (x_1^2 +\cdots +x_4^2)^{1/2}$.
The relevant part of the effective  Euclidean action  is
\beqn
&&\hskip -1cm
S_E =  \int d^4 x \int_0^{\pi R} dy \,
\bigg\{ \frac{1}{2} (\partial_\mu A_y^{13})^2
+\frac{1}{2}(\partial_\mu A_y^{19})^2
  + V_\eff(A_y)  \bigg\} \cr
\noalign{\kern 10pt}
&&\hskip -.3cm
= {4\pi^2 \over g_4^2} \int_0^\infty \rho^3 d\rho ~
\bigg\{ {1\over 2} \Big( {d \phi\over d\rho} \Big)^2
     + U(\phi) \bigg\} ~~,  \cr
\noalign{\kern 15pt}
&&\hskip -1cm
U(\phi) = {1\over 2} \pi R^5 g_4^2 ~
   V_\eff(a,b)\Big|_{a=b=\phi} ~~.
\label{Euclidean1}
\eeqn
The  minima of $U(\phi)$  are located at $\phi=0$ and 1 ($mod$ 2).
It is our disposal to normalize  $U(\phi)$ such that
$U(\phi_{\rm false})=0$.

The Euler-Lagrange equation for $\phi$ is
\beeq
\frac{d^2 \phi}{d\rho^2} + \frac{3}{\rho} \frac{d \phi}{d\rho} =
U'(\phi) ~~,
\label{phiEq1}
\eneq
which is equivalent to the equation of motion for a particle in
a potential $-U(\phi)$ with friction.  The bounce solution needs to
satisfy the boundary conditions
\beqn
&&\hskip -1cm
\lim_{\rho \to \infty} \phi(\rho)  = \phi_{\rm false} ~~, \cr
\noalign{\kern 10pt}
&&\hskip -1cm
\frac{d \phi}{d \rho} \, \Bigg|_{\rho = 0} =0 ~~.
\label{phiEq2}
\eeqn
At $\rho=0$ $\phi$ is at $\phi_0$  near $\phi_{\rm true}$.
It starts to roll down the hill of the
potential $-U(\phi)$, passes the valley, and climbs up  the hill to
reach $\phi_{\rm false}$ at $\rho=\infty$.   The solution can be easily
found numerically by the shooting method, which amounts to finding a right
value of $\phi_0$.

In the non-supersymmetric case $U(\phi)$ is given, from (\ref{Veff8}),  by
\beqn
&&\hskip -1cm
U(\phi) =  {3 g_4^2 \over 16 \pi^6}  ~ \widetilde U_1(\phi) \cr
\noalign{\kern 10pt}
&&\hskip -1cm
\widetilde U_1(\phi) =  -[f_5(\phi) - f_5(1)] -
         {3\over 4} ~ [f_5(2\phi) - f_5(0)]  ~~.
\label{phi-pot1}
\eeqn
$\widetilde U_1(\phi)$ has a global minimum at $\phi=0$,
a local minimum at $\phi=1$, and a maximum at $\phi=0.6124$. Its values
are $\widetilde U_1(0) = - {31\over 16} \zeta_R(5) = -2.00905 $,
$\widetilde U_1(1)=0$, and $\widetilde U_1(0.6124)= 0.73749$.

To estimate the tunneling rate we rescale $\rho$ by
$z = (3 g_4^2/ 16 \pi^6)^{1/2} ~\rho$.  Then the equation to be
solved becomes
\beqn
&&\hskip -1cm
\frac{d^2 \phi}{dz^2} + \frac{3}{z} \frac{d \phi}{dz}
       = \widetilde U_1'(\phi) ~~, \cr
\noalign{\kern 10pt}
&&\hskip -1cm
\phi'(0) = 0 ~~,~~ \phi(\infty) = 1 ~~.
\label{phiEq3}
\eeqn
The Euclidean action for the bounce is
\beeq
S_E =  {64\pi^8 \over 3 g_4^4} \int_0^\infty z^3 dz ~
\bigg\{ {1\over 2} \Big( {d \phi\over dz} \Big)^2
     + \widetilde U_1(\phi) \bigg\} ~~.
\label{Euclidean2}
\eneq
The equation (\ref{phiEq3}) does not contain any parameter.
The thin-wall approximation cannot be applied in the problem.
As the magnitude of $\widetilde U(\phi)$ is $O(1)$, the integral
in (\ref{Euclidean2}) is expected to be $O(1)$.  The detailed
numerical evaluation shows that the integral is about 1.396.
Hence for $g_4^2 /4\pi \sim 1/50$, $S_E \sim 4.5 \times 10^7$. 
The false vacuum is practically stable.  We note that the bounce 
solution starts at $\phi (0) = 3.274 \times 10^{-4}$ and make
transition  in the interval $[1.0, 2.5]$ in $z$.  The friction term in
(\ref{phiEq3}) is very effective.  The size of the bounce is about $143
R$.   

In the supersymmetric case $U(\phi)$ is obtained from
(\ref{SUSYeffpot}).  For $N_h=2$ and $\beta \ll 1$ it is
approximately given by
\beqn
&&\hskip -1cm
U(\phi) \sim  {3 \beta^2 g_4^2 \over 8 \pi^4}  ~ \widetilde U_2(\phi) \cr
\noalign{\kern 10pt}
&&\hskip -1cm
\widetilde U_2(\phi) =  f_3(\phi) - f_3(2 \phi) ~~.
\label{phi-pot2}
\eeqn
This time $\widetilde U_2(\phi)$ has a global minimum at $\phi=1$,
a local minimum at $\phi=0$, and a maximum at $\phi=0.3803$.
Its values are
$\widetilde U_2(1) = - {7\over 4} \zeta_R(3) = -2.1036$,
$\widetilde U_2(0)=0$, and $\widetilde U_2(0.3803)= 0.960244$.
After rescaling $z = (3 \beta^2 g_4^2/8 \pi^4)^{1/2} \rho$,
$\phi$ satisfies the same equation as in (\ref{phiEq3}) where
$\widetilde U_1$ is replaced by $\widetilde U_2$.
The Euclidean action for the bounce is
\beeq
S_E =  {32 \pi^6 \over 3 \beta^2 g_4^4} \int_0^\infty z^3 dz ~
\bigg\{ {1\over 2} \Big( {d \phi\over dz} \Big)^2
     + \widetilde U_2(\phi) \bigg\} ~~.
\label{Euclidean3}
\eneq
One remark is in order.  The potential $\widetilde U_2(\phi)$ 
is not regular at $\phi=1$.  Its second derivative 
$\widetilde U_2''(1)$ diverges.  The bounce solution starts
very close to the true vacuum; $\phi(0) = 1 - 1.0 \times 10^{-7}$.
It stays near $\phi = 1$ for  a while, then make a transition to 
$\phi=0$ in the interval $[1.5, 2.3]$ in $z$.  Again the friction term
in the equation is very effective.   The integral in
(\ref{Euclidean3}) is numerically evaluated to be 1.4.  
$S_E$ is large.   For $\beta = 10^{-2}$ and $g_4^2/4\pi = 1/50$,  for
instance, $S_E \sim 2.3 \times 10^9$.  The size of the bounce
is about $(8\pi^4/3 g_4^2)^{1/2} (R/\beta)$, or
$(M_{\rm SUSY})^{-1}$. 

The life time of the false vacuum is extremely long.  In the SUSY SU(5) model
the false vacuum with the $SU(3) \times SU(2) \times U(1)$
symmetry eventually decays into the true vacuum with the $SU(2) \times
SU(2) \times U(1) \times U(1)$ symmetry.  Its tunneling rate  per
unit time is $\Gamma \sim  1.04 \times 10^{-10^9} $ year$^{-1}$
for
$V = 1.7 \times 10^{78}$m$^3$,
$M=M_{\rm GUT}=10^{16}$GeV,  (or $M_{\rm SUSY}=10^3$GeV)  $\beta= 10^{-13}$,
and
$g_4^2/4\pi = 1/50$. In other words, it is, in practice,  stable.  

The long life time of the false vacuum originates from the form of 
the effective potential $V_\eff$ generated by the dynamics of
Wilson line phases, $\theta$'s. The salient feature of $V_\eff(\theta)$
is that it takes the form $V_\eff(\theta) = k\, R^{-5} \, \widetilde
U(\theta)$ where
$\widetilde U(\theta)$ does not contain any parameters in the theory.
The overall  coefficient $k$ depends on various  parameters in the
theory. In four dimensional (low-energy) spacetime small $k$ translates
to large $S_E$ and longer life time.

\sxn{Summary and discussions}  

We have investigated $SU(5)$ gauge theories on $M^4 \times (S^1/Z_2)$ with
particular
attention on physics of the orbifold boundary condition and the dynamics
of the Wilson lines.
First we have given general arguments for classifying the equivalence classes
of the boundary conditions, evaluating the effective potential for the
left-over Wilson lines, and determining the residual gauge symmetry.
These arguments are applicable to theories with arbitrary gauge groups.
In particular, the structure of the Hosotani mechanism has been
sharpened  in (i) $\sim$ (v) of section 2.5.
Here we summarize it briefly.
`Wilson line phases $\theta$'s along non-contractible loops become
physical degrees of freedom, whose dynamics selects
the physical vacuum  configuration  minimizing
the effective potential $V_\eff$.
If the configuration is nontrivial, the gauge symmetry
is either spontaneously broken or enhanced by radiative corrections and gauge
fields for broken generators and all extra-dimensional
components of gauge fields become massive.
Some of matter fields also acquire masses.
Rearrangement of gauge symmetry takes place.
The physical symmetry of the theory, in general,
differs from the symmetry of the boundary
conditions.  Several sets of  boundary conditions having distinct
symmetry can be related  by   boundary-condition-changing
gauge transformations, thus belonging to the same equivalence class.'
The Hosotani mechanism guarantees the  same physics
in each equivalence class.  The effective potential $V_\eff$ , and so do  
the expectation values of the Wilson line phases.  Physical symmetry of the
theory is determined by the combination of the boundary conditions and the
expectation values of the Wilson line phases.  Theories in the same
equivalence class of the boundary conditions have the same physical symmetry
and physics content.

We have also examined non-SUSY and SUSY $SU(5)$ models on $M^4 \times
(S^1/Z_2)$ to demonstrate  rearrangement of gauge symmetry, showing how the
symmetry is reduced or enhanced by
quantum corrections, depending on the matter content.
In the nonsupersymmetric $SU(5)$ model with the boundary conditions
(\ref{case3})  the minimum of $V_\eff$ is found at
three points,
depending on the particle content as shown in (\ref{minimum1}).
The physical symmetry is $SU(3) \times SU(2) \times U(1)$ when
there are no bulk fermion fields.

The presence of bulk
fermions can lead to the spontaneous breaking of color $SU(3)$.
Systems with various boundary conditions in (\ref{boundary1}) have been
shown to possess the same
physics and to be gauge equivalent to each other due to the Hosotani
mechanism. We have found that zero modes of the extra-dimensional
components, $A_y$,  of gauge fields acquire masses by radiative corrections.
In the supersymmetric model with Scherk-Schwarz SUSY breaking,
color $SU(3)$ can be  spontaneously broken at the global minimum of
$V_\eff$
if there exist more than one Higgs hypermultiplets ($N_h \ge 2$) in the bulk.
The zero modes of $A_y$'s acquire masses of order of SUSY breaking.

It has been also found that the false vacuum appearing in the model
has sufficiently long lifetime, much larger than the age of the universe.
It is due to the special form of the effective potential for the
dimensionless Wilson line phases.

In this paper we have shown how models with simple boundary conditions
have rich structure in the pattern of symmetry breaking/enhancement and mass
generation.   As an interesting subject, it is yet left over to
construct a more realistic grand unified model based on
higher-dimensional space-time in which the Hosotani mechanism and the
orbifold symmetry breaking conspire to reduce symmetries of the system to
those of the standard model.  Implementation of the electroweak symmetry
breaking is also necessary.  Last but not least, our treatment of Higgs
fields in the fundamental representation is incomplete in the sense that
they have not been unified with gauge fields.  With a larger gauge group to
start with, all of the Higgs fields in the standard model can be unified in
a single gauge field multiplet.   We shall come back to these problems in 
near future.

\vskip 1.5cm

\leftline{\bf Acknowledgments}

We would like to thank S.\ Komine,  C.S.\ Lim and K.\ Oda for many 
useful discussions.  M.H.\ is  grateful to  Y.\ Takubo for his
helpful advice in programming.  N.H.\ and Y.H.\ would like to thank 
the Summer Institute 2002 held at Fuji-Yoshida for its hospitality
where a part of the work was carried out.
This work was supported in part by  Scientific Grants
from the Ministry of Education and Science, Grant No.\ 13135215,
Grant No.\ 13640284 (M.H. and Y.H.), Grant No.\ 14039207, 
Grant No.\ 14046208, \ Grant No.\ 14740164 (N.H.).

\newpage

\axn{Residual gauge invariance and mode expansion in $SU(2)$ models}

\vskip -.5cm
It is instructive to classify the residual gauge invariance and mode
expansion with orbifold boundary conditions in $SU(2)$  models.   The
residual gauge invariance is given by (\ref{residual1});
\beqn
&&\hskip -1cm
\Omega(x,y+2\pi R) \,  U  =  U \, \Omega (x,y) \cr
\noalign{\kern 5pt}
&&\hskip -1cm
\Omega(x,-y) \, P_0  = P_0 \,  \Omega (x,y) \cr
\noalign{\kern 5pt}
&&\hskip -1cm
\Omega(x,\pi R -y) \, P_1 = P_1  \Omega (x,\pi R + y)  ~~.
\label{residual3}
\eeqn
$P_0^2 = P_1^2 = I$ and $U= P_1 P_0$.
One can always diagonalize $P_0=P_0^\dagger $ utilizing global $SU(2)$
invariance.

\bigskip

\noindent Case (i) ~ $P_0, P_1= I$ or $-I$

In this case $U=I$ or $-I$ and the conditions in (\ref{residual3}) reduce to
\beeq
\Omega(x,y+2\pi R)   =    \Omega (x,y)  ~~,~~
\Omega(x,-y)   =  \Omega (x,y) ~~.
\label{residual4}
\eneq
There remains the $SU(2)$ gauge invariance.
\beqn
&&\hskip -1cm
\Omega(x,y) = \exp \Big\{ i \sum_{a=1}^3 \omega_a(x,y) \tau_a \Big\} \cr
\noalign{\kern 10pt}
&&\hskip -1cm
\omega_a(x,y) = {1\over \sqrt{\pi R}} ~ \omega_{a,0}(x)
  + \sqrt{{2\over \pi R}} \sum_{n=1}^\infty \omega_{a,n}(x) ~ \cos {ny\over
R} ~~.
\label{gauge-tr1}
\eeqn
$\{ \omega_{a,0}(x); a=1,2,3 \}$ represents the low energy $SU(2)$ gauge
invariance.

Mode expansion in this case is well known.  Each component of fields
is characterized by the values of $(P_0, P_1)$.  A field
$\phi^{(P_0, P_1)}(x,y)$ is expanded as
\beqn
&&\hskip -1cm
\phi^{(++)} (x,y) = {1\over \sqrt{\pi R}} ~ \phi_0 (x)
  + \sqrt{{2\over \pi R}}\sum_{n=1}^\infty \phi_n (x) ~
    \cos {ny\over R}   \cr
\noalign{\kern 10pt}
&&\hskip -1cm
\phi^{(--)} (x,y) = \sqrt{{2\over \pi R}}
   \sum_{n=1}^\infty \phi_n (x) ~ \sin {ny\over R}  \cr
\noalign{\kern 10pt}
&&\hskip -1cm
\phi^{(+-)} (x,y) = \sqrt{{2\over \pi R}}
   \sum_{n=0}^\infty \phi_n (x) ~ \cos {(n+ \onehalf) y\over R}  \cr
\noalign{\kern 10pt}
&&\hskip -1cm
\phi^{(-+)} (x,y) = \sqrt{{2\over \pi R}}
   \sum_{n=0}^\infty \phi_n (x) ~ \sin {(n+ \onehalf) y\over R}  ~~,
\label{expansion5}
\eeqn
where $\pm$ indicates $\pm 1$.

\bigskip

\noindent Case (ii) ~ $P_0= I$ or $-I$, and $P_1 = \tau_3$

If $P_0 \propto I$, $P_1$ can be diagonalized by a global $SU(2)$
rotation.  One can take without loss of generality $P_1 = \tau_3$
if $P_1$ is not proportional to $I$.
In this case $U = i \tau_3$.  The symmetry of the boundary conditions
is $U(1)$.  The conditions in (\ref{residual3}) read
\beeq
\Omega(x,y+2\pi R)  \tau_3 =  \tau_3  \Omega (x,y)  ~~,~~
\Omega(x,-y)   =  \Omega (x,y) ~~.
\label{residual5}
\eneq
The residual gauge invariance is given by
\beqn
&&\hskip -1cm
\pmatrix{\omega_1 (x,y) \cr \omega_2(x,y) \cr}
=  \sqrt{{2\over \pi R}} \sum_{n=0}^\infty
\pmatrix{ \omega_{1,n}(x) \cr \omega_{2,n}(x) \cr}
    ~ \cos {(n+\onehalf) y\over R} \cr
\noalign{\kern 10pt}
&&\hskip -1cm
\omega_3(x,y) = {1\over \sqrt{\pi R}} ~ \omega_{3,0}(x)
  +\sqrt{{2\over \pi R}} \sum_{n=1}^\infty \omega_{3,n}(x)
   ~ \cos {ny\over R} ~~.
\label{gauge-tr2}
\eeqn
where $\omega_a$'s are defined in (\ref{gauge-tr1}).
$\{ \omega_{3,0}(x)\}$ represents the low energy $U(1)$ gauge
invariance.

Mode expansion is the same as in Case (i), and is given by
(\ref{expansion5}).

\bigskip

\noindent Case (iii) ~ $P_0= \tau_3$,
and $P_1 = \tau_3 \, e^{2\pi i(\alpha_1 \tau_1 + \alpha_2 \tau_2)}$

Without loss of generality we set $\alpha_1 = 0$ and $\alpha_2=\alpha$.
$U = e^{-2\pi i\alpha \tau_2}$.  The symmetry of the boundary conditions
is minimal;  there is none.  The conditions in (\ref{residual3}) are
\beqn
&&\hskip -1cm
\Omega(x,y+2\pi R)
   =  e^{-2\pi i\alpha \tau_2} \,  \Omega (x,y) \,
    e^{+2\pi i\alpha \tau_2} ~~,\cr
\noalign{\kern 10pt}
&&\hskip -1cm
\Omega(x,-y)   =  \tau_3 \Omega (x,y) \tau_3 ~~,
\label{residual6}
\eeqn
which read
\beqn
&&\hskip -1cm
\omega_a(x, -y) = \mp \omega_a(x,y) \qquad {\rm for} ~~
\cases{a=1, 2 \cr a=3 ~~.\cr}   \cr
\noalign{\kern 15pt}
&&\hskip -1cm
\omega_2(x, y+ 2\pi R) = \omega_2(x,y) \cr
\noalign{\kern 10pt}
&&\hskip -1.3cm
\pmatrix{\omega_1(x, y+ 2\pi R) \cr \omega_3(x, y+ 2\pi R)\cr}
= \pmatrix{ \cos 4\pi \alpha & \sin 4\pi \alpha \cr
            -\sin 4\pi \alpha & \cos 4\pi \alpha \cr}
\pmatrix{\omega_1(x, y) \cr \omega_3(x, y)\cr} ~~.
\label{residual7}
\eeqn
The residual gauge symmetry is given by
\beqn
&&\hskip -1cm
\omega_2(x,y) = \sqrt{{2\over \pi R}} \sum_{n=1}^\infty
   \omega_{2, n}(x) \sin {ny\over R} \cr
\noalign{\kern 10pt}
&&\hskip -1cm
\pmatrix{\omega_1(x,y) \cr \omega_3(x,y) \cr} =
{1\over \sqrt{\pi R}} \sum_{n=-\infty}^\infty
v_n(x) \pmatrix{ \sin \mybig{(n+2\alpha) y\over R} \cr
                  \cos \mybig {(n+2\alpha) y\over R} \cr}  ~~.
\label{gauge-tr3}
\eeqn
The low energy $U(1)$ gauge invariance appears at
$\alpha= 0, \pm\onehalf, \pm 1 , \cdots$.

Mode expansion depends on representations in $SU(2)$.  For a doublet
field $\phi$, or more specifically for
\beqn
&&\hskip -1cm
\phi(x,-y) = \tau_3 \phi(x,y) ~~, \cr
\noalign{\kern 5pt}
&&\hskip -1cm
\phi(x,y+2\pi R) = e^{-2\pi i\alpha \tau_2} \, \phi(x,y)
= \pmatrix{ \cos 2\pi \alpha & -\sin 2\pi \alpha \cr
            \sin 2\pi \alpha & ~~\cos 2\pi \alpha \cr} \,
       \phi(x,y)
\label{rep1}
\eeqn
the expansion is given by
\beqn
\noalign{\kern 5pt}
&&\hskip -1cm
\pmatrix{\phi_1(x,y) \cr \phi_2(x,y) \cr} =
{1\over \sqrt{\pi R}} \sum_{n=-\infty}^\infty
\phi_n(x) \pmatrix{ \cos \mybig{(n+\alpha) y\over R} \cr
                  \sin \mybig {(n+\alpha) y\over R} \cr}  ~~.
\label{expansion6}
\eeqn
For a triplet field $\phi_a(x,y)$, the expansion is the same as for
$\omega_a(x,y)$ in (\ref{gauge-tr3}).
As the parameter $\alpha$ changes, the mode expansion also changes.
When $\alpha$ shifts to $\alpha +1$, $\phi_n(x)$  of a doublet field
shifts to  $\phi_{n-1}(x)$.  The resultant spectrum returns to the
original one.
 
The expansion (\ref{expansion6}) with the boundary condition
(\ref{rep1}) constitutes the most typical one.  In the computations 
of the effective potential $V_\eff(a,b)$ all fields decompose 
into pairs of the type (\ref{expansion6}).

\axn{Generators and  structure constants of $SU(5)$}

Generators $\Big\{ ~ \onehalf \lambda_a ~; ~a= 1 , \cdots, 24 ~ \Big\}$
of $SU(5)$ are given as follows. $\lambda_3, \lambda_8, \lambda_{9},
\lambda_{12}$ are diagonal matrices given by
\beqn
&&\hskip -1cm
\lambda_3 = {\rm diag~} (1, -1,0,0,0) \cr
\noalign{\kern 12pt}
&&\hskip -1cm
\lambda_8 = {1\over \sqrt{3}} ~ {\rm diag~}  (1, 1,-2,0,0) \cr
\noalign{\kern 5pt}
&&\hskip -1cm
\lambda_9 = {1\over \sqrt{15}} ~{\rm diag~}  (2, 2,2,-3,-3) \cr
\noalign{\kern 12pt}
&&\hskip -1cm
\lambda_{12} = {\rm diag~}  (0, 0,0,1,-1) ~~.
\label{generator1}
\eeqn
Other generators are summarized by the following table;
\begin{eqnarray}
&&\left(
\begin{array}{ccc|cc}
      &(1,2)&(4,5)&(13,14)&(15,16) \\
   (1,2)   &     &(6,7)&(17,18)&(19,20)\\
   (4,5)   &(6,7)    &     &(21,22)&(23,24)\\ \cline{1-5}
   (13,14)   &(17,18)     &  (21,22)   &       &(10,11)  \\
   (15,16)   &(19,20)     & (23,24)    & (10,11)      &
\end{array}
\right) ~~.
\label{table1}
\end{eqnarray}
Here $(1,2)$, for instance,  indicates
\begin{eqnarray}
\lambda_1=\left(
\begin{array}{ccccc}
   & 1 & & &   \\
  1 &  & & &   \\
   &  & & &   \\
   &  & & &   \\
   &  & & &
\end{array}
\right),\;\;\lambda_2=\left(
\begin{array}{ccccc}
   & -i & & &   \\
  i &  & & &   \\
   &  & & &   \\
   &  & & &   \\
   &  & & &
\end{array}
\right)   ~~.
\end{eqnarray}
Generators of $G_\SM = SU(3)\times SU(2)\times U(1)$ are
$\G_\SM = \Big\{ ~ \onehalf \lambda_a ~; ~a= 1 , \cdots, 12 ~ \Big\}$.

In the text and Appendices C and D we need to evaluate eigenvalues of
$D^M(A^0) D_M(A^0)$ or $\gamma^M D_M(A^0)$ where $A^0$ is given by
(\ref{Ay}) and (\ref{Theta2}).  As $A^0$ is non-vanishing only in the
$\lambda_{13}$ and $\lambda_{19}$ components, one needs to know only
the structure constants $f_{abc}$'s with $a$ being either 13 or 19.
($f_{abc}$ is normalized such that
$[\lambda_a, \lambda_b] = 2i f_{abc} \lambda_c$.)
They are given by,
\beqn
&&\hskip -1cm
f_{13,1,18}=f_{13,2,17}=f_{13,3,14}=f_{13,4,22}=f_{13,5,21} \cr
\noalign{\kern 5pt}
&&\hskip .3cm
=f_{13,16,10}=f_{13,11,15}=-f_{13,12,14}=-{1\over 2} ~~, \cr
\noalign{\kern 15pt}
&&\hskip -1cm
f_{19,1,16}=-f_{19,2,15}=-f_{19,3,20}=f_{19,6,24}=f_{19,7,23} \cr
\noalign{\kern 5pt}
&&\hskip .3cm
=-f_{19,10,18}=-f_{19,11,17}=-f_{19,20,12}=-{1\over 2} ~~, \cr
\noalign{\kern 10pt}
&&\hskip -1cm
f_{13,8,14}= f_{19,8,20}=-{1\over 2\sqrt{3}} ~~,~~
f_{13,9,14}=f_{19,9,20}=-{\sqrt{5}\over 2\sqrt{3}} ~~.
\label{structure1}
\eeqn

\axn{Derivation of $V^{g+gh}_\eff (a,b)$} 


In this appendix we derive (\ref{Veff5}) which is the sum of
contributions from the gauge fields and the ghost fields in  the
adjoint representation of $SU(5)$.  We present the more general result
(\ref{Veff-BC4}) for the boundary condition (BC4) in (\ref{boundary1}).
(\ref{Veff5}) corresponds to a special case $p=q=0$ in (BC4).

The evaluation of the effective potential at the one loop level
is reduced to finding the excitation spectrum of fields.  We start
the discussion by examining the spectrum for a pair of fields
$\{ \phi_1(x,y),\phi_2(x,y)\}$ subject to the boundary condition
(\ref{rep1}) whose Lagrangian density is given by
\beeq
{\cal L}_{\rm pair} =
 {1\over 2} \bigg(\dd_y \phi_1 - {\gamma\over R} \phi_2 \bigg)^2
+ {1\over 2} \bigg(\dd_y \phi_2 + {\gamma\over R} \phi_1 \bigg)^2 ~~.
\label{pair1}
\eneq
Note that ${\cal L}_{\rm pair}(x,-y) = {\cal L}_{\rm pair}(x,y)$ and
${\cal L}_{\rm pair}(x,y+2\pi R) = {\cal L}_{\rm pair}(x,y)$. 
Making use of the expansion (\ref{expansion6}), one finds
\beqn
&&\hskip -1cm
\int_0^{\pi R} dy \, {\cal L}_{\rm pair}
= {1\over 4} \int_{-\pi R}^{\pi R} dy \,  (\phi_1, \phi_2) 
\pmatrix{ 
- \dd_y^2 + \myfrac{\gamma^2}{R^2} & \myfrac{2\gamma}{R} \dd_y \cr
-\myfrac{2\gamma}{R} \dd_y & - \dd_y^2 + \myfrac{\gamma^2}{R^2} \cr}
\pmatrix{\phi_1 \cr \phi_2 \cr} \cr
\noalign{\kern 10pt}
&&\hskip -.0cm
= {1\over 2} \sum_{n=-\infty}^\infty 
{(n+\alpha +\gamma)^2\over R^2} \, \phi_n(x)^2 ~~.
\label{pair2}
\eeqn
The Kaluza-Klein excitation spectrum in the fifth dimension is 
$(n+\alpha+\gamma)^2/R^2$ ($-\infty < n < \infty$), which we
symbolically summarize as
\beeq
\big[ ~ \phi_1, \phi_2 ; \alpha, \gamma~ \big]
= \bigg\{  {(n+\alpha +\gamma)^2\over R^2}  ~;
-\infty < n < \infty \bigg\} 
\label{pair3}
\eneq
Here $\alpha$ is the boundary condition parameter in (\ref{rep1})
whereas $\gamma$ represents the amount of mixing caused
by nonvanishing Wilson line phases as in (\ref{pair1}). 
Note that $[ ~ \phi_1, \phi_2 ; \alpha, \gamma~ ] =
[ ~ \phi_1, \phi_2 ; -\alpha, -\gamma~ ]$.

\ignore{
Each component of the fields has a definite $Z_2$ parity, for
example, $A_{\mu}^a(x,y)$ $(a=1,...,12)$  and $A_{y}^a(x,y)$
$(a=13,...,24)$ have the  $Z_2$ even parity and $A_{\mu}^a(x,y)$
$(a=13,...,24)$ and $A_{y}^a(x,y)$ $(a=1,...,12)$ have  the $Z_2$ odd
one in the $Z_2$ parity assignment (\ref{case2}).
}

In (\ref{Veff}), $V_\eff [A^0]^{g+gh}$ is given by 
$-(D-2){i \over 2}\Tr  \ln  D_{L}^{0} D^{0L}$.
The trace of operator $D_{L}^{0} D^{0L}$ is evaluated as the sum of its 
eigenvalues. Hence we must first find the eigenvalues of
$D_{L}^{0} D^{0L} = \partial_\mu \partial^\mu + D_y(A_y^0)D_y(A_y^0)$
for the adjoint representation.
Let us  introduce a field $B =\sum_a B_a {\lambda_a  \over 2}$
in the adjoint representation.
The eigenvalues are found by diagonalizing the bilinear form 
${\rm tr}\, (B  D_y(A^0)D_y(A^0) B)$
in an appropriate  orthogonal basis as given in
(\ref{expansion6}).
Insertion of $A_y^0 ={1 \over 2gR}(a \lambda_{13} + b \lambda_{19})$
leads to 
\begin{eqnarray}
&&\hskip -1cm 
{\cal L} =  - \tr (B D_y(A^0)D_y(A^0) B)
\sim \tr (\partial_y B + ig [A^0, B])^2 \cr
\noalign{\kern 10pt}
&&\hskip -.5cm 
= \tr \left(\sum_c [\partial_y B_c -{1\over R}\sum_{a,b}f_{abc} 
\theta_aB_b]{\lambda_c\over 2}\right)^2   \cr
\noalign{\kern 10pt}
&&\hskip -.5cm 
={1 \over 2}\sum_{c=1}^{24} \left(\partial_y B_c -{a\over 
R}\sum_{d} f_{13,d,c}B_d-{b\over R}
 \sum_{d}f_{19,d,c} B_d \right)^2 ~~.
\label{trace1}
\end{eqnarray}
Making use of  (\ref{structure1}),  (\ref{trace1}) is
written  as 
\beqn
&&\hskip -1cm 
{\cal L} = {\cal L}_1 +  {\cal L}_2 + {\cal L}_4 + {\cal L}_6 ~~, \cr
\noalign{\kern 15pt}
&&\hskip - 1cm 
{\cal L}_1 = 
{1\over 2}\Big\{ (\dd_y B_{13})^2 + (\dd_y B_{19})^2 \Big\} \cr
\noalign{\kern 15pt}
&&\hskip - 1cm 
{\cal L}_2 = {1\over 2}\bigg\{ 
\Big(\dd_y B_4 -{a\over 2R} B_{22}\Big)^2 
       +\Big(\dd_y B_{22} +{a\over  2R} B_4\Big)^2   \cr
\noalign{\kern 5pt}
&&\hskip 0.5cm 
 +\Big(\dd_y B_5 -{a\over 2R} B_{21}\Big)^2
             +\Big(\dd_y B_{21} +{a\over  2R} B_5\Big)^2 \cr
\noalign{\kern 5pt}
&&\hskip 0.5cm 
 +\Big(\dd_y B_6 -{b\over 2R} B_{24}\Big)^2
   +\Big(\dd_y B_{24} +{b\over  2R}  B_6\Big)^2 \cr
\noalign{\kern 5pt}
&&\hskip 0.5cm 
 +\Big(\dd_y B_7 -{b\over 2R} B_{23}\Big)^2
      +\Big(\dd_y B_{23} +{b\over 2R} B_7\Big)^2  \bigg\} ~~, \cr
\noalign{\kern 15pt}
&&\hskip - 1cm 
{\cal L}_4 = {1\over 2}\bigg\{ 
\Big(\dd_y B_1 -{a\over 2R} B_{18}-{b\over 2R} B_{16} \Big)^2
 +\Big(\dd_y B_{10} +{a\over 2R} B_{16}+{b\over 2R} B_{18} \Big)^2\cr
\noalign{\kern 5pt}
&&\hskip 0.5cm 
+ \Big(\dd_y B_{16} -{a\over 2R} B_{10} +{b\over 2R} B_{1} \Big)^2
 + \Big(\dd_y B_{18} +{a\over 2R} B_{1} -{b\over 2R} B_{10} \Big)^2\cr
\noalign{\kern 5pt}
&&\hskip 0.5cm 
+ \Big(\dd_y B_2 -{a\over 2R} B_{17}+{b\over 2R} B_{15} \Big)^2
+ \Big(\dd_y B_{11} -{a\over 2R} B_{15}+{b\over 2R} B_{17} \Big)^2 \cr
\noalign{\kern 5pt}
&&\hskip 0.5cm 
+ \Big(\dd_y B_{15} +{a\over 2R} B_{11} -{b\over 2R} B_{2} \Big)^2
+\Big(\dd_y B_{17} +{a\over 2R} B_{2} -{b\over 2R} B_{11} \Big)^2 
   \bigg\} \cr
\noalign{\kern 15pt}
&&\hskip - 1cm 
{\cal L}_6 = {1\over 2}\bigg\{ 
\Big(\dd_y B_3    -{a\over 2R} B_{14}+{b\over 2R} B_{20} \Big)^2
 +\Big(\dd_y B_{12} +{a\over 2R} B_{14}-{b\over 2R} B_{20} \Big)^2\cr
\noalign{\kern 5pt}
&&\hskip 0.5cm 
 +\Big(\dd_y B_8 -{a\over 2\sqrt{3}R} B_{14}
    -{b\over2\sqrt{3}R} B_{20} \Big)^2
 +\Big (\dd_y B_9 -{\sqrt{5} \,  a\over 2\sqrt{3}R} B_{14}
  - {\sqrt{5} \, b\over2\sqrt{3}R} B_{20} \Big)^2  \cr
\noalign{\kern 5pt}
&&\hskip 0.5cm 
+\Big(\dd_y B_{14}+{a\over 2R} B_{3}-{a\over 2R} B_{12}
  +{a\over 2\sqrt{3}R} B_{8}
  +{\sqrt{5} \, a\over 2\sqrt{3}R} B_{9} \Big)^2  \cr
\noalign{\kern 5pt}
&&\hskip 0.5cm 
 +\Big(\dd_y B_{20}-{b\over 2R} B_{3}+{b\over 2R} B_{12}
 +{b\over 2\sqrt{3}R} B_{8}
  +{\sqrt{5} \,b\over 2\sqrt{3}R} B_{9} \Big)^2
 ~ \bigg\} ~~. 
\label{trace2}
\eeqn

We start to evaluate the contributions from ${\cal L}_2$.  There
are four pairs [4-22], [5-21], [6-24], and [7-23].  They are already in 
the basic form of (\ref{rep1}) and (\ref{pair1}).  For the [4-22] pair
of $A_\mu$, 
\beqn
&&\hskip -1cm
\pmatrix{B_4 \cr B_{22}\cr} (x,-y) 
 = \tau_3 \pmatrix{B_4 \cr B_{22}\cr} (x,y) \cr
\noalign{\kern 10pt}
&&\hskip -1cm
\pmatrix{B_4 \cr B_{22}\cr} (x,y+2\pi R) 
 = e^{i\pi p \tau_2}  \pmatrix{B_4 \cr B_{22}\cr} (x,y) ~~.
\label{pair4}
\eeqn
For the [4-22] pair of $A_y$ the $Z_2$ parity at $y=0$ is reversed
in (\ref{pair4}), while the $S^1$ boundary condition remains the same. 
The same relations hold for the [5-21] pair.   
Similar relations hold for the [6-24] pair where
$(B_4,B_{22}, p)$ are replaced by $(B_6,B_{24}, q)$ in (\ref{pair4}).
To sum up, we have
\beqn
&&\hskip -1cm 
\big[ ~ B_4, B_{22} ; - \onehalf p ,  \onehalf a ~ \big] ~~,~~
\big[ ~  B_5, B_{21} ; - \onehalf p ,  \onehalf a ~ \big] \cr
\noalign{\kern 5pt}
&&\hskip -1cm 
\big[ ~ B_6, B_{24} ; - \onehalf q ,  \onehalf b ~ \big] ~~,~~
\big[ ~  B_7, B_{23} ; - \onehalf q ,  \onehalf b ~ \big]  ~~,
\label{pair5}
\eeqn
for the $A_\mu$ and ghost components.  For the $A_y$ components
$[  B_4, B_{22} ; - \onehalf p ,  \onehalf a ]$, for instance,
is replaced by $[  B_{22}, B_{4} ; + \onehalf p ,  -\onehalf a ]$.
The resulting spectrum is the same as for $A_\mu$. 
Adding the contribution of $\dd_\mu \dd^\mu$ in the lower four 
dimensions, one finds that the ${\cal L}_2$ part yields 
\beeq
2 \sum_{n = -\infty}^{\infty} 
\ln \Bigg\{ -k^2 + 
   \left({n - \onehalf p + \onehalf a \over  R} \right)^2 \Bigg\}
+ 2 \sum_{n = -\infty}^{\infty} 
\ln \Bigg\{ -k^2 + 
   \left({n - \onehalf q + \onehalf b \over  R} \right)^2 \Bigg\} ~.
\label{2entry-a}
\eneq

The part ${\cal L}_4$ in (\ref{trace2}) consists of two sets
[1-10-16-18] and [2-11-15-17].  Both of them decompose to fundamental 
pairs.  Take the set [1-10-16-18], for example.   Introducing
\beeq
C_\pm = {B_{16} \pm B_{18} \over \sqrt{2}} ~~,~~
D_\pm = {B_{1} \pm B_{10} \over \sqrt{2}} ~~,
\label{CDfield1}
\eneq
one finds 
\beqn
&&\hskip -1.5cm 
{\cal L}_4^{\rm  [1-10-16-18]}
 = {1\over 2}\bigg\{ 
\Big(\dd_y D_+ + {a-b \over 2R} C_- \Big)^2
+ \Big(\dd_y C_- - {a-b \over 2R} D_+ \Big)^2 \cr
\noalign{\kern 10pt}
&&\hskip 1.5cm 
+ \Big(\dd_y D_- - {a+b \over 2R} C_+ \Big)^2
+ \Big(\dd_y C_+ + {a+b \over 2R} D_- \Big)^2  \bigg\} ~~.
\label{CDfield2}
\eeqn
The boundary conditions for the $A_\mu$ components are
\beqn
&&\hskip -1.0cm 
\pmatrix{D_\pm \cr C_\mp \cr} (x,-y) 
 = \tau_3 \pmatrix{D_\pm \cr C_\mp \cr} (x,y) \cr
\noalign{\kern 10pt}
&&\hskip -1cm
\pmatrix{D_\pm \cr C_\mp \cr} (x,y+2\pi R) 
 = e^{i\pi (\mp p +q) \tau_2}  
      \pmatrix{D_\pm \cr C_\mp \cr} (x,y) ~~.
\label{pair6}
\eeqn
It follows that 
\beeq
\big[ ~ D_+, C_- ;  \onehalf (p-q) ,  -\onehalf (a-b) ~ \big] ~~,~~
\big[ ~ D_-, C_+ ; - \onehalf (p+q) ,  \onehalf (a+b) ~ \big] 
\label{pair7}
\eneq
for the $A_\mu$ components.   The same spectrum holds for the 
set [2-11-15-17].  The spectrum is unchanged for the $A_y$ components 
as well.  The contributions from ${\cal L}_4$ are 
\beqn
&&\hskip -1cm 
2 \sum_{n = -\infty}^{\infty}  
\ln \Bigg\{ -k^2 
  + {1\over R^2}
\left[ ~ n + \onehalf (p+q) - \onehalf (a+b)  ~ \right]^2
   \Bigg\}    \cr
\noalign{\kern 10pt}
&&\hskip 1.0cm 
+2 \sum_{n = -\infty}^{\infty}  
\ln \Bigg\{ -k^2 + {1\over R^2} 
   \left[ ~ n + \onehalf (p-q) - \onehalf (a-b)  ~ \right]^2
   \Bigg\} ~~ .
\label{4entry}
\eeqn

${\cal L}_6$ in (\ref{trace2}) simplifies by expressing the 
diagonal components $B_3$, $B_8$, $B_9$, and $B_{12}$ in an 
appropriate basis.  It is obvious that one should take, instead of
(\ref{generator1}), 
\beqn
&&\hskip -1cm
\Lambda_1 = \diag (1,0,0, -1,0) \cr
\noalign{\kern 10pt}
&&\hskip -1cm
\Lambda_2 =  \diag  (0, 1,0,0, -1) \cr
\noalign{\kern 10pt}
&&\hskip -1cm
\Lambda_3 = {1\over \sqrt{2}} ~\diag  (1, -1, 0,1,-1) \cr
\noalign{\kern 5pt}
&&\hskip -1cm
\Lambda_4 = {1\over \sqrt{10}} \diag  (1, 1,-4,1,1) 
\label{generator2}
\eeqn
as a basis for diagonal elements.  Accordingly new fields $C_j$'s are
introduced by
\beeq
C_1 \Lambda_1 + C_2 \Lambda_2 + C_3 \Lambda_3 + C_4 \Lambda_4  
= B_3 \lambda_3 + B_8 \lambda_8 + B_9 \lambda_9 + B_{12} \lambda_{12}
~~,
\label{Cfield1}
\eneq
in terms of which ${\cal L}_6$ becomes
\beqn
&&\hskip -1.5cm 
{\cal L}_6
 = {1\over 2}\bigg\{ 
\Big(\dd_y C_1 - {a \over R} B_{14} \Big)^2
+ \Big(\dd_y B_{14} + {a \over R} C_1 \Big)^2 \cr
\noalign{\kern 10pt}
&&\hskip .0cm 
+\Big(\dd_y C_2 - {b \over R} B_{20} \Big)^2
+ \Big(\dd_y B_{20} + {b \over R} C_2 \Big)^2 \cr
\noalign{\kern 10pt}
&&\hskip .0cm 
+(\dd_y C_3 )^2 + (\dd_y C_4 )^2 \bigg\} ~~.
\label{Cfield2}
\eeqn

$(C_1, B_{14})$ and $(C_2, B_{20})$ form pairs.  Their boundary
conditions for $A_\mu$'s are
\beqn
&&\hskip -1.0cm 
\pmatrix{C_1 \cr B_{14} \cr} (x,-y) 
 = \tau_3 \pmatrix{C_1 \cr B_{14} \cr} (x,y) ~~, \cr
\noalign{\kern 10pt}
&&\hskip -1cm
\pmatrix{C_1 \cr B_{14} \cr} (x,y+2\pi R) 
 = e^{2\pi ip  \tau_2}  
      \pmatrix{C_1 \cr B_{14} \cr} (x,y) ~~,
\label{pair8}
\eeqn
and 
\beqn
&&\hskip -1.0cm 
\pmatrix{C_2 \cr B_{20} \cr} (x,-y) 
 = \tau_3 \pmatrix{C_2 \cr B_{20} \cr} (x,y) ~~, \cr
\noalign{\kern 10pt}
&&\hskip -1cm
\pmatrix{C_2 \cr B_{20} \cr} (x,y+2\pi R) 
 = e^{2\pi iq  \tau_2}  
      \pmatrix{C_2 \cr B_{20} \cr} (x,y) ~~,
\label{pair9}
\eeqn
Contributions from $(C_1,C_2,B_{14},B_{20})$ are summarized as
\beeq
\big[ ~ C_1, B_{14} ;  -p ,  a ~ \big] ~~,~~
\big[ ~C_2, B_{20} ; - q,  b ~ \big] 
\label{pair10}
\eneq
so that we have, in the effective potential, 
\beqn
\sum_{n = -\infty}^{\infty} 
\ln \Bigg\{ -k^2 + \left({n + p - a \over R} \right)^2 
\Bigg\}
+ \sum_{n = -\infty}^{\infty} 
\ln \Bigg\{ -k^2 + \left({n + q - b \over  R} \right)^2 
  \Bigg\}  ~~.
\label{6entry}
\eeqn

Finally contributions from $(B_{13}, B_{19})$ in ${\cal L}_1$ 
and $(C_3, C_4)$ in ${\cal L}_6$ combine to result in a simple form.
Their Lagrangian is independent of the Wilson line phases $(a,b)$.
All of these fields are periodic on $S^1$.  
For  $A_\mu$ components, $(B_{13}, B_{19})$ has $Z_2$ parity  $-1$,
whereas  $(C_3, C_4)$ has $+1$.  The sign is reversed for $A_y$.
Hence $(B_{13}, B_{19}, C_3, C_4)$ gives
\beqn
2 \sum_{n = -\infty}^{\infty} 
\ln \Bigg\{ -k^2 +  {n^2 \over R^2} \Bigg\} ~~.
\label{7entry}
\eeqn

Summing  (\ref{2entry-a}),  (\ref{4entry}),
(\ref{6entry}) and (\ref{7entry}) and putting $p=q=0$, 
we have arrived at the expression (\ref{Veff5}).
The effective potential for the general boundary condition (BC4)
is obtained by replacing $(a,b)$ by $(a-p, b-q)$.

\axn{Contributions from fermions in the bulk}

Let us consider the case in which there are fermions in
the ${\bf 5}$ and ${\bf 10}$ representation of $SU(5)$
propagating in the bulk.
They contribute to $V_\eff$.
We shall find that the contribution from the representation ${\bf r}$
is same as that from its conjugate representation $\bar{\bf r}$ as   
seen from the derivation below.

First we study the contribution from fermions $\psi_i(x,y)$, 
$(i = 1, \cdots ,  5)$ in the ${\bf 5}$ representation.
The orbifold boundary conditions are  given by
\beqn
&&\hskip -1cm
\psi(x,-y) = \pm   P_0 \gamma^5 \psi(x,y) \cr
\noalign{\kern 10pt}
&&\hskip -1cm
\psi(x, \pi R -y) = \pm  e^{i\pi\beta} P_1
  \gamma^5 \psi(x,\pi R + y)  \cr
\noalign{\kern 10pt}
&&\hskip -1cm
\psi(x, y+ 2\pi R) =   e^{-i\pi\beta} P_1 P_0 \psi(x,y) 
\label{OrbiBC-F1}
\eeqn
where $\beta$ is $0$ or $1$. The left and right handed components
are defined by 
$\psi_L (x,y) = \onehalf (1 - \gamma _5) \psi (x,y)$ and
$\psi_R (x,y) = \onehalf (1 + \gamma _5) \psi (x,y)$.

Let us consider the case with  the $+$ sign in (\ref{OrbiBC-F1}).  (The
formula for the case of the $-$ sign is obtained by interchanging L and
R.)  For the pair $(\psi_1, \psi_4)$ the condition (\ref{OrbiBC-F1})
reads
\beqn
&&\hskip -1cm
\pmatrix{ \psi_1 \cr i \psi_4 \cr}  (x, -y) 
= - \tau_3 \gamma^5 \pmatrix{ \psi_1 \cr i \psi_4 \cr}  (x, y) \cr
\noalign{\kern 10pt}
&&\hskip -1cm
\pmatrix{ \psi_1 \cr i \psi_4 \cr}  (x, y + 2\pi R) 
= e^{-i\pi (p+\beta) \tau_2}
 \pmatrix{ \psi_1 \cr i \psi_4 \cr}  (x, y) ~~.
\label{OrbiBC-F2}
\eeqn
It follows from (\ref{rep1}) and (\ref{expansion6}) that 
the fields in the pair are expanded as 
\beqn
&&\hskip -1.5cm
\pmatrix{ \psi_1(x, y) \cr i \psi_4(x, y) \cr}
= {1\over \sqrt{\pi R}} \sum_{n=-\infty}^\infty
\Bigg\{ ~
\psi_L^{(n)} (x) 
\pmatrix{ \cos \mybig \Big( n+\onehalf(p+\beta) \Big) { y\over R} \cr
          \noalign{\kern 5pt}
          \sin \mybig \Big (n+\onehalf(p+\beta) \Big) { y\over R} \cr}  \cr
\noalign{\kern 25pt}
&&\hskip +4.5cm
+ ~ \psi_R^{(n)} (x) 
\pmatrix{ \sin \mybig \Big( n-\onehalf(p+\beta) \Big) { y\over R} \cr
          \noalign{\kern 5pt}
          \cos \mybig \Big( n-\onehalf(p+\beta) \Big) { y\over R} \cr}
~ \Bigg\} ~~.
\label{expansion-F1}
\eeqn

In (\ref{Veff}), $V_\eff [A^0]^{\rm fermion}$ is given by 
$-f(D){i \over 2}\Tr  \ln  D_{L}^{0} D^{0L}$.
Eigenvalues of  $D_{L}^{0} D^{0L} = \dd_\mu \dd^\mu - D_y^2$
in the 1-4 subspace  are easily found  in the basis
of $(\psi_1, i\psi_4)$.  Notice that
\beqn
&&\hskip -1.cm
( \psibar_1, \psibar_4) 
(- D_y^2 ) \pmatrix{\psi_1 \cr \psi_4 \cr} \cr
\noalign{\kern 10pt}
&&\hskip -1.cm
= ( \psibar_1, -i \psibar_4)  
\pmatrix{ 
- \dd_y^2 + \myfrac{a^2}{4 R^2}   & - \myfrac{a}{R} \dd_y \cr
 \myfrac{a}{R} \dd_y & - \dd_y^2 + \myfrac{a^2}{4 R^2} \cr}
\pmatrix{\psi_1 \cr i\psi_4 \cr}
\label{fermi-pair1}
\eeqn
which takes the same form as in (\ref{pair2}).  Hence in the notation of
(\ref{pair3}) the spectrum is summarized as
\beeq
\big[ ~ \psi_{1L}, i \psi_{4L} ; \onehalf (p+\beta) , - \onehalf a ~]
~~,~~
\big[ ~  i \psi_{4R}, \psi_{1R} ; -\onehalf (p+\beta) ,  \onehalf a ~] ~.
\label{Fspectrum1}
\eneq
Similarly for the 2-5 components we have
\beeq
\big[ ~ \psi_{2L}, i \psi_{5L} ; \onehalf (q+\beta) , - \onehalf b ~]
~~,~~
\big[ ~  i \psi_{5R}, \psi_{2R} ; -\onehalf (q+\beta) ,  \onehalf b ~] ~.
\label{Fspectrum2}
\eneq
The contribution from $\psi_3$ does not contain any dependence on
$a$ or $b$.  The spectrum is given by $(n+\onehalf \beta)^2/R^2$
($n$: integers) after combining contributions from the L and R 
components.  With all these contributions added together each  fermion 
multiplet in {\bf 5} yields, for $\Tr \ln D_M^0 D^{0M}$, 
\beqn
&&\hskip -1.cm
 \sum_{n = -\infty}^{\infty}  \ln \left\{ -k^2 
 + \left( {n + \onehalf \beta \over R}  \right)^2 \right\}  \cr
\noalign{\kern 10pt}
&&\hskip -1cm 
 + 2 \sum_{n = -\infty}^{\infty}  \ln \left\{ - k^2 
   + \left( {n - \onehalf (a-p-\beta) \over R}\right)^2 \right\}  \cr
\noalign{\kern 10pt}
&&\hskip -1cm 
 + 2 \sum_{n = -\infty}^{\infty}  \ln \left\{ - k^2 
   + \left( {n - \onehalf (b-q-\beta) \over R}\right)^2 \right\} 
~~ .
\label{Fentry1}
\eeqn

Next we study the contribution from fermion $\psi_{ij} = - \psi_{ji}$, 
$(i,j = 1, \cdots ,  5)$ in the ${\bf 10}$ representation.
The $Z_2$ transformation property is given by
\beqn
&&\hskip -1cm
\psi_{ij}(x,-y) = \pm   {(P_0)_i}^{i'} {(P_0)_j}^{j'} \gamma^5 
\psi_{i'j'}(x,y)  ~~, \cr
\noalign{\kern 10pt}
&&\hskip -1cm
\psi_{ij}(x, \pi R -y) 
= \pm  e^{i\pi\beta} {(P_1)_i}^{i'} {(P_1)_j}^{j'}
  \gamma^5 \psi_{i'j'}(x,\pi R + y) ~~,  \cr
\noalign{\kern 10pt}
&&\hskip -1cm
\psi_{ij}(x,y+ 2 \pi R) = e^{-i\pi\beta}
  {(P_1 P_0)_i}^{i'} {(P_1 P_0)_j}^{j'}
  \psi_{i'j'}(x,y)
\label{OrbiBC-F3}
\eeqn
where $\beta$ is either $0$ or $1$.
We consider the case with the $+$ sign in (D.8).

The covariant derivative for $\psi_{ij}$ is given by
\begin{equation}
({D_M \psi})_{jk} \equiv
     \partial_M \psi_{jk} 
+ ig \Big\{ {(A_M)_j}^l \psi_{lk} + {(A_M)_k}^l \psi_{jl} \Big\}   ~~.
\end{equation}
The kinetic term is decomposed as
\beqn
&&\hskip -1cm
{1\over 2} \psibar {}^{ij} i\gamma^5 (D_y \psi)_{ij}
= {\cal L}_1 + {\cal L}_2 + {\cal L}_4 ~~, \cr
\noalign{\kern 10pt}
&&\hskip -1cm
{\cal L}_1 = \psibar_{14} i\gamma^5 \dd_y \psi_{14}
+ \psibar_{25} i\gamma^5 \dd_y \psi_{25} \cr
\noalign{\kern 15pt}
&&\hskip -1cm
{\cal L}_2 =
(\psibar_{34} , \psibar_{13}) \,  i\gamma^5
\pmatrix{ \dd_y  & -\myfrac{ia}{2R}  \cr
         -\myfrac{ia}{2R} & \dd_y \cr}
      \pmatrix{ \psi_{34} \cr \psi_{13} \cr} \cr
\noalign{\kern 5pt}
&&\hskip 2.5cm
+ (\psibar_{35} , \psibar_{23}) \,  i\gamma^5
\pmatrix{ \dd_y  & -\myfrac{ib}{2R}  \cr
         -\myfrac{ib}{2R} & \dd_y \cr}
      \pmatrix{ \psi_{35} \cr \psi_{23} \cr} ~~, \cr
\noalign{\kern 15pt}
&&\hskip -1cm
{\cal L}_4
=(\psibar_{12}, \psibar_{15} , \psibar_{24},  \psibar_{45} ) \, i\gamma^5
\pmatrix{
\dd_y  & \myfrac{ib}{2R} & -\myfrac{ia}{2R} & 0 \cr
\myfrac{ib}{2R} & \dd_y & 0 & \myfrac{ia}{2R} \cr
-\myfrac{ia}{2R} & 0 & \dd_y & -\myfrac{ib}{2R} \cr
 0 & \myfrac{ia}{2R} & -\myfrac{ib}{2R} & \dd_y  \cr}
 \pmatrix{  \psi_{12} \cr \psi_{15} \cr \psi_{24} \cr \psi_{45} \cr} ~~.
\label{Fdecomposition1}
\eeqn

The condition (D.8) for the pair  $(\psi_{34},\psi_{13})$  reads 
\beqn
&&\hskip -1cm
\pmatrix{ \psi_{34} \cr i \psi_{13} \cr}  (x, -y) 
= - \tau_3 \gamma^5 \pmatrix{ \psi_{34} \cr i \psi_{13} \cr}  (x, y) \cr
\noalign{\kern 10pt}
&&\hskip -1cm
\pmatrix{ \psi_{34} \cr i \psi_{13} \cr}  (x, y + 2\pi R) 
= e^{i\pi (p+\beta) \tau_2}
 \pmatrix{ \psi_{34} \cr i \psi_{13} \cr}  (x, y) ~~,
\label{OrbiBC-F4}
\eeqn
which has the same form as (\ref{OrbiBC-F2}) when 
 $(\psi_1,\psi_4,p)$ is replaced by $(\psi_{34},\psi_{13},-p)$.
As
\beqn
&&\hskip -1.cm
( \psibar_{34}, \psibar_{13}) 
(- D_y^2 ) \pmatrix{\psi_{34} \cr \psi_{13} \cr} \cr
\noalign{\kern 10pt}
&&\hskip -1.cm
=(\psibar_{34}, -i \psibar_{13})  
\pmatrix{ 
- \dd_y^2 + \myfrac{a^2}{4 R^2}   &  \myfrac{a}{R} \dd_y \cr
- \myfrac{a}{R} \dd_y & - \dd_y^2 + \myfrac{a^2}{4 R^2} \cr}
\pmatrix{\psi_{34} \cr i\psi_{13} \cr} ~~,
\label{fermipair2}
\eeqn
the spectrum is given by
\beeq
\big[ ~ \psi_{34L}, i \psi_{13L} ; - \onehalf (p+\beta) ,  \onehalf a ~]
~~,~~
\big[ ~  i \psi_{13R}, \psi_{34R} ; \onehalf (p+\beta) , - \onehalf a ~] ~.
\label{Fspectrum3}
\eneq
Similarly we have for the pair  $(\psi_{35},\psi_{23})$
\beeq
\big[ ~ \psi_{35L}, i \psi_{23L} ; -\onehalf (q+\beta) ,  \onehalf b ~]
~~,~~
\big[ ~  i \psi_{23R}, \psi_{35R} ; \onehalf (q+\beta) , - \onehalf b ~] ~.
\label{Fspectrum4}
\eneq

${\cal L}_4$ in (\ref{Fdecomposition1}) is simplified when expressed in
terms of
\beeq
E_\pm = {\psi_{12} \pm \psi_{45} \over \sqrt{2}} ~~,~~
F_\pm = {\psi_{15} \pm \psi_{24} \over \sqrt{2}} ~~.
\label{EFfield1}
\eneq
One finds from (\ref{OrbiBC-F3}) that
\beqn
&&\hskip -1cm
\pmatrix{i F_- \cr E_+ \cr}  (x, -y) 
= - \tau_3 \gamma^5 \pmatrix{ i F_- \cr E_+  \cr}  (x, y) \cr
\noalign{\kern 10pt}
&&\hskip -1cm
\pmatrix{i F_-  \cr E_+ \cr}  (x, y + 2\pi R) 
= e^{i\pi (p+q+\beta) \tau_2}
 \pmatrix{ i F_- \cr E_+ \cr}  (x, y) ~~.
\label{OrbiBC-F5}
\eeqn
As
\beqn
&&\hskip -1.cm
( \overline{F}_-, \overline{E}_+) 
(- D_y^2 ) \pmatrix{ F_- \cr E_+ \cr} \cr
\noalign{\kern 10pt}
&&\hskip -1.cm
=(-i \overline{F}_- , \overline{E}_+ ) 
\pmatrix{ 
-\dd_y^2+\bigg( \myfrac{a+b}{2R} \bigg)^2 & \myfrac{a+b}{R}\dd_y \cr
-\myfrac{a+b}{R}\dd_y & -\dd_y^2 +\bigg(\myfrac{a+b}{2R}\bigg)^2 \cr}
\pmatrix{i F_- \cr E_+ \cr} ~~, 
\label{fermi-pair3}
\eeqn
the spectrum is given by
\beeq
\big[ ~ i F_{-L}, E_{+L} ; -\onehalf (p+q+\beta),\onehalf (a+b) ~ \big]
~~,~~
\big[ ~ E_{+R} ,i F_{-R}  ; \onehalf(p+q+\beta), -\onehalf(a+b) ~ \big] ~.
\label{Fspectrum5}
\eneq
Relations for the pair $(E_-,F_+)$ are obtained by replacing
$(E_+,F_- ; p+q,a+b)$ in (\ref{OrbiBC-F5}) and (\ref{fermi-pair3}) by
$(E_-,F_+ ; -p+q, -a+b)$.  The spectrum is given by
\beeq
\big[ ~ i F_{+L}, E_{-L} ; \onehalf(p-q+\beta)  , - \onehalf (a-b) ~ \big] 
~~,~~
\big[ ~ E_{-R} ,i F_{+R}  ;-\onehalf(p-q+\beta),   \onehalf(a-b)  ~ \big]
~~.
\label{Fspectrum6}
\eneq
The contributions from  $(\psi_{14},\psi_{25})$ depend on 
neither Wilson line phases $a,b$ nor boundary condition  parameters $p,q$. 
Their spectrum is given by $(n+\onehalf \beta)^2/R^2$ ($n$: integers).
 With all these contributions added each fermion 
multiplet in {\bf 10} yields, for $\Tr \ln D_M^0 D^{0M}$, 
\beqn
&&\hskip -1cm
 2 \sum_{n = -\infty}^{\infty} \ln \left\{ -k^2 
   + \left( {n+\onehalf  \beta \over R}\right)^2 \right\}  \cr
\noalign{\kern 10pt}
&&\hskip -1cm
+ 2 \sum_{n = -\infty}^{\infty} \ln \left\{ -k^2 
   + \left({n - \onehalf ( a-p-\beta)} \over R \right)^2 \right\} \cr
\noalign{\kern 10pt}
&&\hskip -1cm
+ 2 \sum_{n = -\infty}^{\infty} \ln \left\{ -k^2 
   + \left({n - \onehalf(  b-q-\beta)}\over R \right)^2 \right\}  \cr
\noalign{\kern 10pt}
&&\hskip -1cm
+ 2 \sum_{n = -\infty}^{\infty} \ln \left\{ -k^2 
    + \left({n -\onehalf( a+b-(p+q)-\beta)} \over R \right)^2 \right\} \cr
\noalign{\kern 10pt}
&&\hskip -1cm
+ 2 \sum_{n = -\infty}^{\infty} \ln \left\{ -k^2 
  + \left({n -\onehalf( a-b-(p-q)-\beta)} \over R \right)^2 \right\} ~~ .
\label{Fentry2}
\eeqn
(\ref{Fentry1}) and (\ref{Fentry2}), after being integrated over $k$, lead 
to (\ref{Veff7}) for $p=q=0$.  

The results in the previous and this appendices show that the effective
potential $V_\eff^{(p,q)} (a,b)$ with the boundary condition (BC4) in 
(\ref{boundary1}) is a function of $a-p$ and $b-q$,  thus establishing 
(\ref{Veff-BC4}).

\vskip 1.5cm

\def\jnl#1#2#3#4{{#1}{\bf #2} (#4) #3}

\def\Zphys{{\em Z.\ Phys.} }
\def\jssc{{\em J.\ Solid State Chem.\ }}
\def\jpsJ{{\em J.\ Phys.\ Soc.\ Japan }}
\def\ptps{{\em Prog.\ Theoret.\ Phys.\ Suppl.\ }}
\def\PTP{{\em Prog.\ Theoret.\ Phys.\  }}

\def\JMP{{\em J. Math.\ Phys.} }
\def\NPB{{\em Nucl.\ Phys.} B}
\def\NP{{\em Nucl.\ Phys.} }
\def\PLB{{\em Phys.\ Lett.} B}
\def\PL{{\em Phys.\ Lett.} }
\def\PRL{\em Phys.\ Rev.\ Lett. }
\def\PRB{{\em Phys.\ Rev.} B}
\def\PRD{{\em Phys.\ Rev.} D}
\def\PRe{{\em Phys.\ Rep.} }
\def\AP{{\em Ann.\ Phys.\ (N.Y.)} }
\def\RMP{{\
em Rev.\ Mod.\ Phys.} }
\def\ZPC{{\em Z.\ Phys.} C}
\def\SCI{\em Science}
\def\CMP{\em Comm.\ Math.\ Phys. }
\def\MPLA{{\em Mod.\ Phys.\ Lett.} A}
\def\IJMPA{{\em Int.\ J.\ Mod.\ Phys.} A}
\def\IJMPB{{\em Int.\ J.\ Mod.\ Phys.} B}
\def\EPJC{{\em Eur.\ Phys.\ J.} C}
\def\PR{{\em Phys.\ Rev.} }
\def\JHEP{{\em JHEP} }
\def\cmp{{\em Com.\ Math.\ Phys.}}
\def\JPA{{\em J.\  Phys.} A}
\def\CQG{\em Class.\ Quant.\ Grav. }
\def\ATMP{{\em Adv.\ Theoret.\ Math.\ Phys.} }
\def\ibid{{\em ibid.} }

\leftline{\bf References}

\renewenvironment{thebibliography}[1]
         {\begin{list}{[$\,$\arabic{enumi}$\,$]}  
         {\usecounter{enumi}\setlength{\parsep}{0pt}
          \setlength{\itemsep}{0pt}  \renewcommand{\baselinestretch}{1.2}
          \settowidth
         {\labelwidth}{#1 ~ ~}\sloppy}}{\end{list}}

\end{document}